\documentclass[12pt]{iopart}

\usepackage{amsfonts}
\usepackage{graphicx}

\usepackage{mathrsfs} 
\usepackage{latexsym} 
\usepackage{amssymb}

\usepackage{amsthm}

\newcommand{\calC}{\mathcal{C}}

\newcommand{\calE}{\mathcal{E}}

\newcommand{\calL}{\mathcal{L}}
\newcommand{\calM}{\mathcal{M}}

\newcommand{\bfR}{\mathbb{R}}
\newcommand{\bfS}{\mathbb{S}}
\newcommand{\bfZ}{\mathbb{Z}}

\newcommand{\bfX}{\mathbb{X}}

\newcommand{\lie}{\mathscr{L}}

\newcommand{\pr}{\mathrm{pr}}

\newcommand{\dual}[1]{\left<{#1}\right>}

\newcommand{\im}{\mathrm{Im}\,}
\renewcommand{\Im}{\im}

\newcommand{\dd}[1]{\frac{\d}{\d #1}}

\newcommand{\dde}{\dd{\epsilon}}
\newcommand{\pdd}[1]{\frac{\partial}{\partial #1}}

\newcommand{\contraction}{\vrule height 0pt depth 0.4pt width 3pt
  \vrule height 7pt depth 0.4pt \kern 3pt}

\newcommand{\be}{\begin{equation}}
\newcommand{\ee}{\end{equation}}

\newtheoremstyle{jvk-thm} %
  {}{}{\itshape}{}{\bfseries}{.}{ }{}
\newtheoremstyle{jvk-rem} %
  {}{}{\upshape}{}{\bfseries}{.}{ }{}

\theoremstyle{jvk-thm}
\newtheorem{definition}{Definition}[section]
\newtheorem{lemma}[definition]{Lemma}
\newtheorem{theorem}[definition]{Theorem}
\newtheorem{prop}[definition]{Proposition}

\newenvironment{mproof}{\textbf{Proof:}\,}{\hfill$\Box$}

\theoremstyle{jvk-rem}
\newtheorem{example}[definition]{Example}
\newtheorem{remarkth}[definition]{Remark}
\newenvironment{remark}{\begin{remarkth}}{\hfill$\diamond$\end{remarkth}}

\newenvironment{AlphaList}{%
  \renewcommand{\theenumi}{\alph{enumi}}%
  \begin{enumerate}}{\end{enumerate}}

\newcommand{\gothso}{\mathfrak{so}}

\newcommand{\gothg}{\mathfrak{g}}

\renewcommand{\d}{\mathrm{d}}

\newcommand{\qfor}{\quad \mbox{for }}
\newcommand{\qforall}{\quad \mbox{for all}\:}

\newcommand{\where}{\quad \mathrm{where}\:}
\newcommand{\qqand}{\quad \mathrm{and} \quad}

\renewcommand{\vec}[1]{\mathbf{#1}}
\newcommand{\url}[1]{\texttt{#1}}

\begin{document}


\title{A class of nonholonomic kinematic constraints in elasticity}
\author{Joris Vankerschaver\footnote{Research Assistant of the
    Research Foundation -- Flanders (FWO-Vlaanderen).}}
\address{Department of Mathematical Physics and Astronomy, Ghent
  University, Krijgslaan 281, B-9000 Ghent, Belgium}

\ead{Joris.Vankerschaver@UGent.be}

\begin{abstract}
  We propose a first example of a simple classical field theory with
  nonholonomic constraints.  Our model is a straightforward
  modification of a Cosserat rod. Based on a mechanical analogy, we
  argue that the constraint forces should be modeled in a special way,
  and we show how such a procedure can be naturally implemented in the
  framework of geometric field theory.  Finally, we derive the
  equations of motion and we propose a geometric integration scheme
  for the dynamics of a simplified model.
\end{abstract}

\maketitle


\section{Introduction}

\paragraph{Motivation}
Nonholonomic constraints in mechanical systems have a long and
illustrious history going back to the work of Hertz at the end of the
nineteenth century.  For nonholonomic constraints in \emph{field
  theories}, the case is not so clear, and to the best of our
knowledge, no convincing example has been proposed to this day.  In
this paper, we intend to rectify this omission by giving a simple
example of a continuum theory with nonholonomic constraints.  The
basic model is that of a Cosserat rod, a special kind of continuum
theory.  This rod moves in a horizontal plane which is supposed to be
sufficiently rough, so that the rod rolls without sliding.

In the Cosserat theory one assumes that the laminae at right angles
to the centerline are rigid discs.  The nonholonomic Cosserat rod
therefore touches the plane along a curve (rather than in an open
subset of the plane, which would be the case for a fully
three-dimensional continuum) and the nonholonomic constraint
translates to the fact that the instantaneous velocity of these
contact points is zero.

\begin{figure}
\begin{center}
  \includegraphics{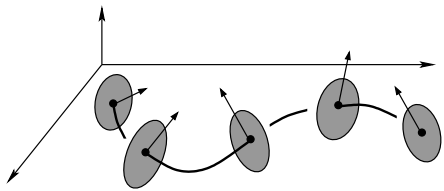}
  \caption{Geometry of the constrained rod}
  \label{fig:discrete}
\end{center}
\end{figure}

Our example can also be modeled as the continuum limit of a
nonholonomic mechanical system.  Consider $N$ rigid discs rolling
vertically without sliding on a horizontal plane, and assume that
these discs are interconnected by flexible beams of length $\ell/N$,
as in figure~\ref{fig:discrete}.  Now let the number of discs go to
infinity, while keeping the total length $\ell$ fixed: the result is
the nonholonomic Cosserat rod.

This mechanical model is interesting for a number of reasons.  First
of all, the nonholonomic field equations are derived by varying the
action with respect to \emph{admissible variations}, and this
obviously requires the specification of a bundle of admissible
variations, or equivalently, a bundle of reaction forces.  In
mechanics, this is commonly done by taking recourse to the principle
of d'Alembert, which states that the virtual work of the reaction
forces is zero.  In field theory, this principle can be interpreted in
a number of non-equivalent ways, and it is the mechanical model which
will eventually determine our choice.  

Secondly, our model is a counterexample to the often-held belief that
constraints in classical field theories are necessarily vakonomic.  In
sections~\ref{sec:rel} and \ref{sec:constraint} these two aspects are
treated more in detail.

\paragraph{Plan of the paper}

After giving a quick overview of jet bundle theory, we derive the
Euler-Lagrange equations in the presence of nonholonomic constraints.
Our treatment relies on the fact that the space of independent
variables is a product of space and time, and that the fields are
sections of a trivial fibre bundle.  This is the case, for instance,
for nonrelativistic elasticity.  These assumptions allow us to split
the jet bundle in a part involving spatial derivatives, and a part
involving derivatives with respect to time.  Using this natural
splitting, we propose a bundle of reaction forces, based on the
mechanical model (to be outlined in section~\ref{sec:constraint}).  As
we shall see, these reaction forces are very similar to the ones used
in mechanics.

The remainder of the paper is then devoted to the study of a specific
example of a nonholonomic field theory.  First, we give an outline of
the theory of Cosserat rods moving in the plane, and we pay particular
attention to aspects of symmetry.  In the second part, we then outline
a suitable class of nonholonomic constraints, and we derive the
equations of motion.  The paper concludes with a brief foray into the
field of geometric integration, where a simple explicit algorithm for
the integration of the nonholonomic dynamics is proposed.

\subsection{Relation with other approaches} \label{sec:rel}

In a number of papers \cite{franca1, franca2}, Bibbona,
Fatibene, and Francaviglia contrasted the vakonomic and the
nonholonomic treatments for classical field theories, and showed that
for relativistic hydrodynamics only the former gives correct results.
Another typical example of a vakonomic constraint is the
incompressibility constraint in nonrelativistic fluid dynamics,
treated by Marsden et al. \cite{MPSW01}.  Many more can be found in
Antman's book \cite{Antman05} and in the papers by Garc\'{\i}a et
al. \cite{Garcia}.

In contrast, our field theory arises as the continuum limit of the
vertically rolling disc, a textbook example of a nonholonomic
mechanical system.  These nonholonomic constraints survive in the
continuum limit and hence provide a very strong motivation for the
study of nonholonomic techniques in field theories.

In previous papers (see \cite{nhfields02, nhfields05, Krupkova05,
  KrupkovaVolny}) various theoretical frameworks were established for
the study of nonholonomic field theories.  Our model fits into these
descriptions, but involves a number of additional ingredients which
cannot be derived from these theoretic considerations alone.  In
particular the bundle of reaction forces takes a special form,
motivated by similar definitions from mechanics.

It should also be noted that similar theories as ours were explored
before by Vignolo and Bruno (see \cite{iperideal02}).  They considered
constraints depending only on the time derivatives of the fields, and
their resulting analysis is therefore more direct.  However, the
underlying philosophy is the same: the constraints are ``(\dots)
purely kinetic restrictions imposed separately on each point of the
continuum''.

\subsection{Modeling the constraint forces} \label{sec:constraint}

\paragraph{Nonholonomic mechanical systems}
The mechanical background is not essential for the description of the
continuum theory, but rather serves as a justification for some of our
definitions.  In particular, it provides a number of valuable clues
regarding the type of constraint forces needed to maintain such a
nonholonomic constraint.  Let $S$ be the configuration space of the
vertically rolling disc, so that the configuration space for the
entire model, consisting of $N$ discs, is the product space $S^N$.
Denote by $\varphi^\alpha_{(i)}$ the constraints of rolling without
sliding imposed on the $i$th wheel; $\varphi^\alpha_{(i)}$ is a
function on $T S^N$.

With these conventions, a \emph{motion} of the system is a curve $t
\mapsto c(t)$ in $S^N$, and a variation of such a motion $c$ is
then a vector field $(X_1, X_2, \ldots, X_N)$ on $S^N$ along $c$,
i.e. a collection of maps $X_i : \bfR \rightarrow TS$ such that
$X_i(t) \in T_{c^i(t)}S$ for all $i = 1, \ldots, N$, where $c^i :=
\pr_i \circ c$.

Let us now consider the one-forms $\Phi_{(i)}^\alpha := J^\ast (\d
\varphi_{(i)}^\alpha)$, where $J$ is the vertical endomorphism on
$TS^N$.  In geometric mechanics, linear combinations of these
one-forms represent the possible reaction forces at the $i$th wheel;
the bundle $F$, defined as
\[
  F := \dual{\Phi^\alpha_{(1)}} \oplus \dual{\Phi^\alpha_{(2)}} \oplus
  \cdots \oplus \dual{\Phi^\alpha_{(N)}}
\]
then represents the totality of all reaction forces along the rod.  In
coordinates, the one-forms $\Phi^\alpha_{(i)}$ are given by
\begin{equation} \label{generators} \Phi^\alpha_{(i)} = \frac{\partial
    \varphi^\alpha_{(i)}}{\partial \dot{y}_{(i)}} \d y_{(i)} \quad
  \left(= \sum_a \frac{\partial \varphi^\alpha_{(i)}}{\partial
      \dot{y}^a_{(i)}} \d y^a_{(i)} \right) \qforall i = 1, \ldots, N.
\end{equation}
Here, $(y_{(i)}, \dot{y}_{(i)})$ is a coordinate system on the $i$th
factor of $TS^N$.  Note that there is no summation over the index
$i$ in (\ref{generators}), and that the summation over individual
coordinates is implicit, as shown in the term between brackets.

Knowing the precise form of the bundle of reaction forces $F$ is
important because the nonholonomic equations of motion are derived by
varying the action with respect to admissible variations.  Moreover,
the principle of d'Alembert shows us that a variation is admissible if
it belongs to the annihilator of $F$, i.e. 
a variation $(X_1, \ldots, X_N)$ of $c$ is admissible if
\begin{equation} \label{alembert}
  \left< \bar{X}_i(t), \alpha(c(t)) \right> = 0, \qforall (i, t) \in 
    \{1, \ldots, N\} \times \bfR \qqand \alpha \in F,
\end{equation}
where $\bar{X}_i$ is a lift of $X_i$ to $T(TS)$ such that $T\tau_S
\circ \bar{X}_i = X_i$.  Note that there is again no summation over $i$.  In
coordinates, this is equivalent to
\begin{equation} \label{abert:coordinates}
  v_i \frac{\partial \varphi_{(i)}}{\partial \dot{y}_{(i)}} \qforall i =
    1, \ldots, N,
\end{equation}
where we have written $X_i = v_i \frac{\partial}{\partial y_{(i)}}$.

In the next paragraph, we will let the number $N$ go to infinity,
while keeping the length $\ell$ constant.  The result is a field
theory, and a reaction force will be a continuous assignment of a
one-form on $TS$ to each point of the centerline of the rod.  This
definition will be the starting point for our treatment in the main
body of the text; once we know the bundle of reaction forces, we can
then derive the nonholonomic field equations.

\paragraph{The continuum model}
In the continuum limit, a field is a map $\phi$ from $[0, \ell] \times
\bfR$ to $S$.  It is customary in classical field theory to view
these fields as sections of a trivial bundle $\pi$, whose base space
is $[0, \ell] \times \bfR$, and with standard fibre $S$.  The role
of the velocity space is then played by the first jet bundle $J^1\pi$,
and the constraints $\varphi_{(i)}$ from the previous paragraph are
replaced by a constraint function $\varphi^\alpha$ on $J^1\pi$.

In field theory, a variation of a field $\phi$ is now a map $X : [0,
\ell] \times \bfR \rightarrow T S$ with the property that $X(s, t) \in
T_{\phi(s, t)} S$; in other words, a vector field along $\phi$.
Taking our cue from (\ref{abert:coordinates}), we say that a variation
$X$ is admissible if the following holds (in coordinates):
\[
  X(s, t)^a \frac{\partial \varphi^\alpha}{\partial y^a_0} = 0,
\]
where we have written $X(s, t) = X(s, t)^a \frac{\partial}{\partial
  y^a}$.  

This condition can be rewritten in intrinsic form by using the
following observation: as we shall show below, there exists a natural
isomorphism between the first jet bundle and the product bundle $\bfR
\times [J^1(M, S) \times_S TS]$.  Now, let $J$ be the vertical
endomorphism on $TS$.  This map has a trivial extension to the whole
of $\bfR \times [J^1(M, S) \times_S TS]$, and by using the natural
isomorphism with $J^1\pi$, we obtain a map $J^\ast$ from
$T^\ast(J^1\pi)$ to itself.  The bundle $F$ of constraint forces is
then generated by the forms $\Phi^\alpha := J^\ast( \d
\varphi^\alpha)$.  The similarity with the mechanical case is obvious.

\section{Lagrangian field theories}

In this paper, we will mostly be concerned with the description of
elastic bodies.  The geometric description of these theories is well
known and we refer to \cite{handbook, Marcelo} for more information. 

Let $M$ be a smooth $n$-dimensional compact manifold, and let $S$ be a
general smooth $m$-dimensional manifold, with $n \le m$.  The points
of $M$ are ``material points'', labelling the points of the body,
whereas $S$ is the physical space in which the body moves.  In most
cases, $S$ will be the Euclidian space $\bfR^3$, whereas $M$ can be
one-, two-, or three-dimensional, corresponding to models of rods,
shells, and three-dimensional continua.  Furthermore, $M$ is assumed
to be oriented, with volume form $\eta_M$.  

On $M$ we consider a coordinate system $(x^i)$ , $i = 1, \ldots, n$,
such that $\eta_M$ can locally be written as $\eta_M = \d x^1 \wedge
\cdots \wedge \d x^n$, and on $S$ we take a coordinate system $(y^a)$,
$a = 1, \ldots, m$.

\subsection{The bundle picture}

In this section, we give a brief overview of the theory of jet
bundles.  For an introduction to classical field theory using jet
bundles, we refer to \cite{CrampinMS, gimmsyI, overview02} and the
references therein.

Consider the fibre bundle $\pi : Y \rightarrow X$, where $X := \bfR
\times M$, $Y := X \times S$, and the projection $\pi$ is the
projection onto the first factor.  The manifold $X$ is equipped with a
coordinate system denoted by $(x^\mu)$, where $\mu = 0, \ldots, n$,
and such that $x^0 := t$.  Similarly, $S$ has a coordinate system
$(x^\mu, y^a)$ which is adapted to the projection in the sense that
the projection $\pi$ is locally given by $\pr_2: (x^\mu, y^a) \mapsto
(x^\mu)$.  Note that $M$ is equipped with a volume form $\eta = \d t
\wedge \eta_M$, which we write in coordinates as $\d^{n+1}x := \d{x}^0
\wedge \cdots \wedge \d{x}^n$.  We will employ the following
short-hand notation:
\[
  \d^n x_\mu := \pdd{x^\mu} \contraction \eta = 
   (-1)^{\mu} \d x_0 \wedge \cdots \wedge \d x^{\mu - 1} \wedge
     \d x^{\mu + 1} \wedge \cdots \wedge \d x^n.
\]

The first jet bundle $J^1\pi$ is the appropriate stage for Lagrangian
first-order field theories.  Its elements are equivalence classes of
sections of $\pi$, where two sections are said to be equivalent at a
point $x \in X$ if they have the same value at $x$ and if their
first-order Taylor expansions at $x$ agree.  The equivalence class of
a section $\phi$ at $x$ is denoted by $j^1_x \phi$.  Hence, $J^1\pi$
is naturally equipped with a projection $\pi_{1, 0}: J^1\pi
\rightarrow Y$, defined by $\pi_{1, 0}(j^1_x \phi) = \phi(x)$, and a
projection $\pi_1: J^1\pi \rightarrow X$, defined by $\pi_1(j^1_x
\phi) = x$.  The induced coordinate system on $J^1\pi$ is written as
$(x^\mu, y^a; y^a_\mu)$.  

Furthermore, we define the manifold $J^1(M, S)$ of jets of mappings
from $M$ to $S$ as the first jet manifold of the trivial bundle
$\pr_1: M \times S \rightarrow M$.  The usual jet bundle projections
$\pi_1$ and $\pi_{1, 0}$ induce projections $\pi_M$, onto $M$, and
$\pi_S$, onto $S$, respectively.

Because of the special structure of the bundle $\pi$, viz. the fact
that $X$ is the product $\bfR \times M$ and that $\pi$ is trivial,
$J^1\pi$ can be written in a special form.  Recall that the fibre
coordinates of $J^1\pi$ represent the derivatives of the fields with
respect to space and time: the decomposition of
lemma~\ref{lemma:bundle} then provides an invariant way of making the
distinction between time derivatives and spatial derivatives.  Such an
invariant decomposition is not possible for general jet bundles.

The bundle $\bfR \times [J^1(M, S) \times_S TS]$, a fibered product,
consists of triples $(t, \kappa, v)$ such that $\pi_S(\kappa) =
\tau(v)$, where $\tau: TS \rightarrow S$ is the tangent bundle
projection.  It is equipped with a projection $\hat{\pi}$ onto $Y$
defined as $\hat{\pi}(t, \kappa, v) = (t, \pi_M(\kappa); \tau(v))$.
Moreover, $\hat{\pi}: \bfR \times [J^1(M, S) \times_S TS] \rightarrow
Y$ is an affine bundle.

\begin{lemma} \label{lemma:bundle} The first jet bundle $J^1\pi$ is
  isomorphic, as an affine bundle over $Y = \bfR \times M \times S$, to
  $\bfR \times [J^1(M, S) \times_S TS]$.
\end{lemma}
\begin{mproof} A more general statement can be found in section~6B of
  \cite{gimmsyII}.

Take any point $(t, m, s)$ in $\bfR \times M \times S$ and consider a
$1$-jet $\gamma$ such that $\pi_{1, 0}(\gamma) = (t, m, s)$.  An
alternative interpretation of $\gamma$ is that of  a linear map $\gamma :
T_{(t, m)}(\bfR \times M) \rightarrow T_s S$.  Consider now the map
$\Psi_{(t, m, s)}$, mapping $\gamma$ to the element of $\bfR \times
[J^1(M, S) \times_S TS]$ given by
\[
\Psi_{(t, m, s)}(\gamma) = \left(t, \gamma(0_t, \cdot), \gamma\left(
    \frac{\partial}{\partial t} \Big|_{t}, 0_m \right)\right),
\]
where $0_m$ and $0_t$ are the zero vectors in $T_m M$ and in $T_t \bfR$,
respectively.  It is easy to check that the map $\Psi$ is an
isomorphism of affine bundles.
\end{mproof}

In coordinates $(y^a, \dot{y}^a)$ on $TS$ and $(x^i, y^a; y^a_{;i})$
on $J^1(M, S)$, the isomorphism of lemma~\ref{lemma:bundle} is given
by $(x^\mu, y^a; y^a_\mu) \mapsto (t; x^i, y^a, y^a_{;i} = y^a_i; y^a,
\dot{y}^a = y^a_0)$.  

For future reference, we remark here that the tangent bundle $TS$ is
equipped with a $1$-$1$ tensor field, denoted by $J: T(TS) \rightarrow
T(TS)$, and given in coordinates by 
\[
  J = \pdd{\dot{y}^a} \otimes \d y^a.
\]
An intrinsic definition can be found in \cite{CrampinTQ}.  The adjoint
of $J$ will be denoted by $J^\ast$ and is a map from $T^\ast(TS)$ to
itself defined by $\dual{J^\ast(\alpha), v} = \dual{\alpha, J(v)}$,
for all $v \in T(TS)$.

In section~\ref{sec:cosserat}, we will encounter higher-order field
theories, in particular of order $2$.  In order to be able to deal
with this type of field theory, we introduce the manifold $J^k\pi$ of
$k$th order jets.  The elements of $J^k\pi$ are again equivalence
classes of sections of $\pi$, where two sections are equivalent at $x
\in X$ if they have the same value at $x$ and if their Taylor
expansions at $x$ agree up to the $k$th order.  The $k$th order jet
bundle is equipped with a number of projections $\pi_{k, l} : J^k \pi
\rightarrow J^l \pi$ (where $l \le k$), constructed by ``truncating''
to order $l$ the Taylor expansion defining an element of $J^k \pi$.  A
detailed account of jet bundles is provided in \cite{Saunders89}.

As a matter of fact, we will only need the third-order jet manifold
$J^3\pi$.  A natural coordinate system on $J^3\pi$ is given by
$(x^\mu, y^a; y^a_\mu; y^a_{\mu\nu}; y^a_{\mu\nu\kappa})$, for $a = 1,
\ldots, m$ and $\mu, \nu, \kappa = 0, \ldots, n$, with the convention
that 
\[
y^a_{\mu\nu} = y^a_{\nu\mu} \qqand y^a_{\mu \nu \kappa} =
y^a_{\sigma(\mu \nu \kappa)}
\] 
for any permutation $\sigma$ of the three indices (expressing the
commutativity of partial derivatives).

\subsection{Covariant field theories of first and second
  order} \label{sec:covft} 

\subsubsection{First-order field theories} \label{sec:first}

The geometry of first-order field theories has been studied by many
authors (see \cite{MPSW01, gimmsyI, overview02, Saunders89, bsf88} and
the references therein for a non-exhaustive survey) and is by now well
established.  In this section, we recall some basic constructions.

Consider a \emph{first-order Lagrangian} $L: J^1\pi \rightarrow \bfR$.
There exists an $(n + 1)$-form $\Theta_L$ on $J^1\pi$, called the
\emph{Cartan form}.  Different intrinsic constructions of $\Theta_L$
can be found in \cite{CrampinMS, gimmsyI, Saunders89}, but here its
coordinate expression will suffice:
\begin{equation} \label{theta}
  \Theta_L = L \d^{n + 1}x + \frac{\partial L}{\partial y^a_\mu} (\d
  y^a - y^a_\nu \d x^\nu) \wedge \d^n x_\mu.
\end{equation}
Let us also define the $(n + 2)$-form $\Omega_L := -\d\Theta_L$ on
$J^1\pi$.

Let $S$ be the \emph{action functional} defined as 
\begin{equation} \label{firstaction}
  S(\phi) = \int_U L(j^1\phi) \eta,
\end{equation}
for each section $\phi$ of $\pi$ with compact support $U$.  We now
look for critical points of this functional under arbitrary
variations, which are defined as follows.

\begin{definition} \label{def:infvar}
  An \emph{infinitesimal variation} of a field $\phi$ defined on $U$
  is a vertical vector field $V$ defined in a neighbourhood of
  $\phi(U)$ in $Y$, with the added restriction that $V(y) = 0$ for all
  $y \in \phi(\partial U)$.
\end{definition}

An infinitesimal variation gives rise to a local one-parameter group
of diffeomorphisms $\Phi_\epsilon: Y \rightarrow Y$ defined in a
neighbourhood of $\phi(U)$.  The fact that $V$ vanishes on
$\phi(\partial U)$ implies that $\Phi_\epsilon(y) = y$ for all $y \in
\phi(\partial U)$.  The composition of $\Phi_\epsilon$ with a section
$\phi$ of $\pi$ is hence a new section of $\pi$, denoted by
$\phi_\epsilon$.

The critical points of $S$ therefore satisfy
\begin{equation} \label{varieren}
  0 = \dde S(j^1 (\Phi_\epsilon \circ \phi) )\Big|_{\epsilon = 0} =
    \int_U \left( \frac{\partial L}{\partial y^a} - \frac{\d}{\d
        x^\mu} \frac{\partial L}{\partial y^a_\mu} \right) V^a
    \d^{n+1}x. 
\end{equation}
As the variations $V$ are arbitrary, we obtain the usual
Euler-Lagrange equations for a section $\phi$ of $\pi$:
\[
\left[ \frac{\partial L}{\partial y^a} - \frac{\d}{\d x^\mu}
  \frac{\partial L}{\partial y^a_\mu} \right](j^2\phi) = 0.
\]

These partial differential equations can be rewritten in intrinsic
form by means of the Cartan form (see \cite{CrampinMS, gimmsyI,
  bsf88}):

\begin{theorem} \label{thm:EL1}
  A section $\phi$ of $\pi$ is a critical point of the action $S$, or,
  equivalently, satisfies the Euler-Lagrange equations, if and only if
  \begin{equation} \label{1stEL}
    (j^1 \phi)^\ast (i_W \Omega_L) = 0
  \end{equation}
  for all vector fields $W$ on $J^1\pi$.
\end{theorem}

\paragraph{Symmetries and Noether's theorem}

Let $G$ be a Lie group acting on $X$ by diffeomorphisms, and on $Y$ by
bundle automorphisms.  For $g \in G$, consider the bundle automorphism
$\Phi_g : Y \rightarrow Y$ with base map $f_g: X \rightarrow X$.  The
prolongation to $J^1\pi$ of the action of $G$ is then defined in terms
of bundle automorphisms $j^1\Phi_g : J^1\pi \rightarrow J^1\pi$,
defined by
\[
  j^1\Phi_g (j^1_x \phi) = j^1_{f_g(x)}( \Phi_g \circ \phi \circ f_g^{-1}).
\]

Consider an element $\xi$ of $\gothg$ and denote the infinitesimal
generator of the prolonged action corresponding to $\xi$ by
$\xi_{J^1\pi}$.  Note that $\xi_{J^1\pi}$ is just $j^1\xi_Y$, the
prolongation of the infinitesimal generator on $Y$ corresponding to
$\xi$.  We recall that if $\xi_Y$ is given in coordinates by
\[
\xi_Y = \xi^\mu
\frac{\partial}{\partial x^\mu} + \xi^a \frac{\partial}{\partial y^a}, 
\]
then its prolongation $j^1\xi_Y$ is defined as
\[
 j^1\xi_Y = \xi^\mu\frac{\partial}{\partial x^\mu} + \xi^a
 \frac{\partial}{\partial y^a} + \left( \frac{\d \xi^a}{\d x^\mu} -
   y^a_\nu \frac{\d \xi^\nu}{\d x^\mu}\right) \frac{\partial}{\partial
   y^a_\mu}. 
\]

We say that a Lagrangian $L$ is \emph{invariant} under the prolonged
action of $G$ if $L \circ j^1 \Phi_g = (\det [Df_g])^{-1} L$ for all
$g \in G$ and $\gamma \in J^1\pi$, where $\det [Df_g]$ is the Jacobian
of $f_g$.  This condition can be expressed concisely as $(j^1
\Phi_g)^\ast (L \eta) = L\eta$.  It can be shown that invariance of
the Lagrangian implies invariance of the Cartan $(n + 1)$-form,
expressed as $j^1\Phi_g^\ast \Theta_L = \Theta_L$ for all $g \in G$,
or, infinitesimally,
\begin{equation} \label{invariance}
  \lie_{\xi_{J^1\pi}} \Theta_L = 0.
\end{equation}

Let $L$ be a $G$-invariant Lagrangian.  Associated to this symmetry is
a map defined as $J^L_\xi := \xi_{J^1\pi} \contraction \Theta_\calL$.
We now define the \emph{momentum map} $J^L : J^1 \pi \rightarrow
\Omega^n(J^1\pi) \otimes \gothg^\ast$ by $\dual{J^L, \xi} = J^L_\xi$.
The importance of the momentum map lies in the following theorem,
which we have taken here from \cite[thm.~4.7]{gimmsyI}:
\begin{prop}[Noether] \label{prop:noether} Let $L$ be an invariant
  Lagrangian.  For all $\xi \in \gothg$, the following conservation
  law holds:
  \[
  \d [(j^1\phi)^\ast J^L_\xi] = 0,
  \]
  for all sections $\phi$ of $\pi$ that are solutions of the
  Euler-Lagrange equations (\ref{1stEL}).
\end{prop}

A comprehensive account of symmetries in classical field theory can be
found in \cite{symm04}.

\subsubsection{Second-order field theories} \label{sec:2ndorder}

Many of the Lagrangians arising in elasticity are of higher order.  In
particular, we will encounter a second-order model in
section~\ref{sec:cosserat}.  In some papers (see \cite{Saunders89,
  Kouranbaeva} and the references therein) a geometric framework for
second-order field theories has been developed and we now recall a
number of relevant results.

A second-order Lagrangian is a function $L$ on $J^2\pi$.  The
corresponding second-order Cartan $(n + 1)$-form is a form on
$J^3\pi$, whose coordinate expression reads
\begin{eqnarray} \label{2ndcartan}
  \Theta_L = \left[ \frac{\partial L}{\partial y^a_\nu} - \frac{\d}{\d
      x^\mu} \left( \frac{\partial L}{\partial y^a_{\nu\mu}} \right)
  \right] \d y^a\wedge \d^n x_\nu + \frac{\partial L}{\partial
    y^a_{\nu\mu}} \d y^a_\nu \wedge \d^n x_\mu \nonumber \\
   +  \left[ L - \frac{\partial L}{\partial y^a_\nu} y^a_\nu +
    \frac{\d}{\d x^\mu} \left( \frac{\partial L}{\partial
        y^a_{\nu\mu}} \right) y^a_\nu - \frac{\partial L}{\partial
      y^a_{\nu\mu}} y^a_{\nu\mu} \right] \d^{n + 1}x.
\end{eqnarray}

Many results from the previous section on first-order field theories
carry over immediately to the higher-order case.  The action $S$ is
defined as
\[
  S(\phi) = \int_U L(j^2\phi) \eta, 
\]
where $\phi$ is again a section of $\pi$ with compact support $U$.  A
section $\phi$ is a critical point of this functional if and only if
it satisfies the second-order Euler-Lagrange equations:
\begin{equation} \label{2ndEL} \left[\frac{\partial L}{\partial
    y^a} - \frac{\d}{\d x^\mu} \left( \frac{\partial
      L}{\partial y^a_\mu} \right) + \frac{\d^2}{\d x^\mu \d
    x^\nu} \left( \frac{\partial L}{\partial y^a_{\mu\nu}}
  \right)\right](j^4\phi) = 0.
\end{equation}
There also exists an intrinsic formulation of the Euler-Lagrange
equations.  We quote from \cite{Kouranbaeva}:

\begin{prop} \label{prop:EL2}
  Let $L$ be a second-order Lagrangian.  A section $\phi$ of $\pi$ is
  a solution of the second-order Euler-Lagrange equations if and only
  if $(j^3\phi)^\ast( W \contraction \Omega_L ) = 0$ for all vector
  fields $W$ on $J^3\pi$.  Here, $\Omega_L := - \d \Theta_L$ is the
  second-order Poincar\'e-Cartan form.
\end{prop}

\begin{remark}
It should be noted that there always exists a Cartan form for
higher-order field theories, but that uniqueness is not guaranteed
(contrary to the first-order case).  However, by imposing additional
conditions, Saunders~\cite{Saunders89} was able to prove uniqueness
for second-order field theories.  This unique form, given in
(\ref{2ndcartan}), was derived by Kouranbaeva and
Shkoller~\cite{Kouranbaeva} by means of a variational argument.
\end{remark}

The action of a Lie group $G$ acting on $Y$ by bundle automorphisms
gives rise to a prolonged action on $J^2 \pi$.  If a Lagrangian is
$G$-invariant with respect to this action, then the momentum map
$J^L \in \Omega^n(J^3\pi) \otimes \gothg^\ast$, defined as $\dual{J^L,
  \xi} = J^L_\xi$, where $J^L_\xi = \xi_{J^3\pi} \Theta_L$, gives rise
to a conservation law: $\d[(j^3\phi)^\ast J^L_\xi] = 0$ for all
sections $\phi$ of $\pi$ that are solutions of the Euler-Lagrange
equations (\ref{2ndEL}).

\subsection{Nonholonomic field theories} \label{sec:nonholft}

\subsubsection{The field equations} \label{sec:fe}

We now derive the Euler-Lagrange equations in the presence of
nonholonomic constraints.  The nonholonomic problem involves the
specification of two distinct elements: the constraint manifold
$\calC$, and the bundle of reaction forces $F$.  Following Marle
\cite{marle98}, we impose no a priori relation between $\calC$ and
$F$.  

The \emph{constraint manifold $\calC$} is a submanifold of $J^1\pi$ of
codimension $k$ and represents the external constraints imposed on the
system.  For the sake of definiteness, we will assume that $\calC$
projects onto the whole of $Y$ (i.e. $\pi_{1, 0}(\calC) = Y$), and
that the restriction $(\pi_{1, 0})_{|\calC}: \calC \rightarrow Y$ is a
fibre bundle.  This need not be an affine subbundle of $J^1\pi$.  For
the benefit of clarity, $\calC$ is assumed to be given here by the
vanishing of $k$ functionally independent functions $\varphi^\alpha$ on
$J^1\pi$:
\[
  \calC := \{ \gamma \in J^1\pi: \varphi^\alpha(\gamma) = 0 \qfor
  \alpha = 1, \ldots, k \}.
\]
The treatment can be easily extended to the case where the
$\varphi^\alpha$s are only locally defined.

Secondly, we assume the existence of a $k$-dimensional codistribution
$F$ on $J^1\pi$, along $\calC$, of \emph{reaction forces}.  The
elements of $F$ are maps $\alpha: \calC \rightarrow T^\ast S$ such
that $\alpha(\gamma) \in T^\ast_s S$, where $s = (\pr_2 \circ \pi_{1,
  0})(\gamma)$.  If we denote by $\pi_{TS}: J^1\pi \rightarrow TS$ the
composition $\pi_{TS} := \pr_3 \circ \Psi$, where $\Psi$ is the
isomorphism defined in lemma~\ref{lemma:bundle}, then the elements of
$F$ can equivalently be viewed as one-forms along the projection
$\pi_{TS}$.  By pull-back, these one-forms then induce proper
one-forms defined along $\calC$.  In local coordinates, an element
$\alpha$ of $F$ can be represented as $\alpha = A_a(x^\mu, y^a,
y^a_\mu) \d y^a$, where the $A_a$ are local functions on $\calC$.

We define the annihilator $F^\circ$ of $F$ as the following subbundle
of $T J^1\pi$ along $\calC$: for all $\gamma \in \calC$, 
\[
  F^\circ_\gamma := \{ v_\gamma \in T_\gamma J^1\pi : 
    \dual{\alpha_\gamma, v_\gamma} = 0 \qforall \alpha_\gamma
  \in F_\gamma\}.
\] 
An arbitrary element $v_\gamma$ of $F^\circ_\gamma$ has the following
form:
\[
  v_\gamma = v^\mu \pdd{x^\mu} + v^a \pdd{y^a} + v^a_\mu \pdd{y^a_\mu},
  \where v^a A^\alpha_a(\gamma) = 0.
\]
Here, we have chosen a basis of sections $A^\alpha_a \d{y^a}$ of $F$.
No further restrictions are imposed on the coefficients $v^\mu$ and
$v^a_\mu$, but note that this is not the end of the story; see the
appendix. 

The local work done by a ``force'' $\alpha$ along a variation $V$ (see
definition~\ref{def:infvar}) is then given by the pairing
$\dual{\alpha(\gamma), j^1V(\gamma)}$, where $\gamma \in \Im j^1\phi$,
and the global work done at time $t$ by the integral $\int_M
(j^1\phi_t)^\ast \dual{\alpha, j^1V} \eta_M$, where $\phi_t$ is the
instantaneous configuration defined by $\phi_t(u) := \phi(t, u)$.

\begin{definition}
  A variation $V$ of a field $\phi$ defined over an open subset $U$
  with compact closure is \emph{admissible} if $(j^1\phi)^\ast (j^1 V
  \contraction \alpha) = 0$ for all $\alpha \in F$.
\end{definition}

\begin{definition}
  A local section $\phi$ of $\pi$, defined on an open subset $U
  \subset X$ with compact closure, is a \emph{solution} of the
  nonholonomic problem determined by $L$, $\calC$, and $F$ if
  $j^1\phi(U) \subset \calC$ and (\ref{varieren}) holds for all
  admissible variations $V$ of $\phi$.
\end{definition}

It follows from (\ref{varieren}) that a local section $\phi$ is a
solution of the nonholonomic problem if it satisfies the
\emph{nonholonomic Euler-Lagrange equations}:
\begin{equation} \label{eq:nh}
  \left[ \frac{\partial L}{\partial y^a} - \frac{\d}{\d x^\mu}
    \frac{\partial L}{\partial y^a_\mu} \right](j^2\phi) =
  \lambda_\alpha A^\alpha_a(j^1\phi) \qqand \varphi^\alpha(j^1\phi) = 0.
\end{equation}
Here, $\lambda_\alpha$ are unknown Lagrange multipliers, to be
determined from the constraints.  This is proved below.

\begin{theorem} \label{thm:EL}
  Let $\phi$ be a section of $\pi$. If $\im j^1\phi \subset \calC$, then
  the following assertions are equivalent:
  \begin{AlphaList}
    \item $\phi$ is a stationary point of the action (\ref{firstaction})
      under admissible variations; \label{eq1}
    \item $\phi$ satisfies the Euler-Lagrange equations
      (\ref{eq:nh}); \label{eq2} 
    \item for all $\pi$-vertical vector fields $V$ such that $j^1_\gamma V
      \in F^\circ_\gamma$ for all $\gamma \in \calC$, \label{eq3}
      \begin{equation} \label{eq:intrinsic}
        (j^1\phi)^\ast (j^1V \contraction \Omega_L) = 0.
      \end{equation}
  \end{AlphaList}
\end{theorem}
\begin{mproof}
Let us first prove the equivalence of (\ref{eq1}) and (\ref{eq3}).
For arbitrary, not necessarily admissible variations, the following
result holds (this is equation~3C.5 in \cite{gimmsyI}):
\[
  \frac{\d}{\d\epsilon} S(\phi_\epsilon)\Big|_{\epsilon = 0}
    = -\int_X (j^1 \phi)^\ast(j_1 V \contraction \Omega_L).
\]
For admissible variations, we have therefore
\[
  \int_X (j^1 \phi)^\ast(j_1 V \contraction \Omega_L) = 0.
\]
Now, we may multiply $V$ by an arbitrary function on $X$ and this
result will still hold true.  The fundamental lemma of the calculus of
variations therefore shows that 
\begin{equation} \label{eqIntrNH}
  (j^1 \phi)^\ast(j_1 V \contraction \Omega_L) = 0,
\end{equation}
for all admissible variations $V$.  By using a partition of unity as
in \cite{gimmsyI}, it can then be shown that (\ref{eq:intrinsic})
holds for all $\pi$-vertical vector fields $V$ such that $\dual{j^1 V,
  \alpha} = 0$ for all $\alpha \in F$.

The equivalence of (\ref{eq2}) and (\ref{eq3}) is just a matter of
writing out the definitions.  In
coordinates, the left-hand side of (\ref{eq:intrinsic}) reads
\[
(j^1 \phi)^\ast(j^1V \contraction \Omega_L) = 
V^a(j^1\phi)\left( \frac{\partial L}{\partial
    y^a}(j^1\phi) - \frac{\d}{\d x^\mu} \frac{\partial L}{\partial
    y^a_\mu}(j^1\phi) \right) \d^{n+1}x,
\]
and this holds for all variations $j^1V \in F^\circ$.  Therefore, if $\phi$
satisfies (\ref{eq:intrinsic}), then there exist $k$ functions
$\lambda_\alpha$ such that
\[
   \frac{\partial L}{\partial
    y^a}(j^1\phi) - \frac{\d}{\d x^\mu} \frac{\partial L}{\partial
    y^a_\mu}(j^1\phi) = \lambda_\alpha A^\alpha_a(j^1\phi).
\]
The converse is similar.
\end{mproof}

\begin{remark}
  In (\ref{eq:intrinsic}), only prolongations of vertical vector
  fields were considered, whereas in similar expressions in
  theorem~\ref{thm:EL1} and proposition~\ref{prop:EL2}, arbitrary
  vector fields occurred.  This is due to the fact, also mentioned in
  the appendix, that only vertical variations are considered.

  For the derivation of the nonholonomic Euler-Lagrange equations, it
  is enough to consider only vertical variations.  However, if one
  wants to prove the nonholonomic Noether theorem for symmetries that
  act nontrivially on the base space, one needs an expression like
  (\ref{eq:intrinsic}) but with $j^1V$ replaced by a vector
  field $W$ which is not necessarily the prolongation of a vertical
  vector field.  As we point out in the appendix, this can be done,
  but then one needs to modify the definition of $F$.
\end{remark}

\subsubsection{The Chetaev principle}

In section~\ref{sec:fe}, we defined reaction forces as certain maps
from $\calC$ to $T^\ast S$, but their exact nature was left
unspecified.  We now conclude our derivation of the nonholonomic field
equations by proposing a concrete definition for these reaction
forces.  As will become clear in a moment, this definition is formally
identical to the one used in mechanics; roughly speaking, the reaction
forces are constructed by composing $\d \varphi^\alpha$ with the
vertical endomorphism $J$ on $TS$, which is what one might call the
Chetaev principle.

Indeed, the vertical endomorphism $J$ on $TS$ trivially extends to a
$(1, 1)$-tensor $\hat{J}$ on $\bfR \times [J^1(M, S) \times_S TS]$,
defined as
\begin{equation}
  \hat{J}(\alpha, \beta, \gamma) := (0, 0, J(\gamma)),
\end{equation}
where $\alpha \in T^\ast_t \bfR$, $\beta \in T^\ast_u J^1(M, S)$, and
$\gamma \in T^\ast_v(TS)$ (and where $(t, u, v) \in \bfR \times
[J^1(M, S) \times_S TS]$).  We denote the adjoint of this map as
$\hat{J}^\ast$.

Let $\varphi^\alpha$ be the $k$ constraint functions.  By means of the
isomorphism $\Psi$ of lemma~\ref{lemma:bundle}, these functions induce
$k$ functions on $\bfR \times [J^1(M, S) \times_S TS]$, which we also
denote by $\varphi^\alpha$.

\begin{definition} \label{def:reaction}
  The \emph{bundle of reaction forces $F$} is the co-distribution on
  $J^1\pi$ locally generated by the following forms: $F =
  \mathrm{Span}(\Phi^\alpha)$, where
  \[
    \Phi^\alpha := \Psi^\ast [ \hat{J}^\ast( \d \varphi^\alpha) ].
  \]
\end{definition}

In local coordinates on $J^1\pi$, the generating forms $\Phi^\alpha$
are given by
\begin{equation} \label{constrforms}
  \Phi^\alpha = \frac{\partial \varphi^\alpha}{\partial y^a_0}
  \d{y^a}. 
\end{equation}
This corresponds to the coordinate expressions based on the mechanical
analogue: compare, for instance, with (\ref{generators}).  Again, we
emphasize that there is an obvious distinction between spatial
derivatives and derivatives with respect to time.

Using these reaction forces, the nonholonomic Euler-Lagrange equations
become 
\[
\left[ \frac{\partial L}{\partial y^a} - \frac{\d}{\d x^\mu}
  \frac{\partial L}{\partial y^a_\mu} \right](j^2\phi) =
\lambda_\alpha \frac{\partial \varphi^\alpha}{\partial y^a_0}(j^1\phi)
\qqand \varphi^\alpha(j^1\phi) = 0,
\]
where the $\lambda_\alpha$ are again a set of unknown Lagrange
multipliers, to be determined from the constraints.

\section{A Cosserat-type model} \label{sec:cosserat}

The theory of \emph{Cosserat rods} constitutes an approximation to the
full three-dimensional theory of elastic deformations of rod-like
bodies.  Originally conceived at the beginning of the twentieth
century by the Cosserat brothers, it laid dormant for more than fifty
years until it was revived by the pioneers of rational mechanics (see
\cite[\S98]{handbook} for an overview of its history).  It is now an
important part of modern nonlinear elasticity and its developments are
treated in great detail for instance in \cite{Antman05}, which we
follow here.

A Cosserat rod can be visualised as specified by a curve $s \mapsto
\mathbf{r}(s)$ in $\bfR^3$, called the \emph{centerline}, to which is
attached a frame $\{\vec{d}_1(s), \vec{d}_2(s), \vec{d}_3(s)\}$,
called the \emph{director frame} (models with different numbers of
directors are also possible).  The rough idea is that the centerline
characterizes the configuration of the rod when its thickness is
neglected, whereas the directors model the configuration of the laminae
transverse to the centerline.  In the Cosserat theory, the laminae are
assumed to deform homogeneously, and therefore the specification of a
director frame in $\bfR^3$, fixed to a lamina, completely specifies the
configuration of that lamina.

In the special case where the laminae are rigid discs at right angles
to the centerline, one can choose the director frame $\{\vec{d}_1,
\vec{d}_2, \vec{d}_3\}$ to be orthogonal with, in addition,
$\vec{d}_2$ and $\vec{d}_3$ of unit length (attached to the laminae)
and $\vec{d}_1$ aligned with $\vec{r}'(s)$, the tangent vector to the
centerline.  If, in addition, the centerline is assumed to be
inextensible, so that we may choose the parameter $s$ to be arclength,
$\vec{d}_1$ is also of unit length and the director frame is
orthonormal.  In this case, the specification of, say, $\vec{d}_2$ is
enough to determine a director frame: putting $\vec{d}_1 \equiv
\vec{r}'$, we then know that $\vec{d}_3 = \vec{d}_1 \times
\vec{d}_2$.  Here and in the following, a prime ($'$) denotes
derivation with respect to $s$.

Here, we will consider the case of a Cosserat rod with an inextensible
centerline and rigid laminae.  In addition, we will assume that the
centerline is planar in the $(x, y)$-plane, which will allow us to
eliminate the director frame almost completely.  The result is a
Lagrangian field theory of second order, to which the results of
section~\ref{sec:covft} can be applied.

\subsection{The planar Cosserat rod}

\begin{figure}
\begin{center}
  \includegraphics{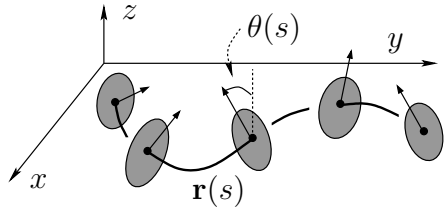}
  \caption{Geometry of the constrained rod}
  \label{fig:rod}
\end{center}
\end{figure}

Consider an inextensible Cosserat rod of length $\ell$ equipped with
three directors.  If we denote the centerline at time $t$ as $s
\mapsto \vec{r}(t, s)$, inextensibility allows us to assume that the
parameter $s$ is the arclength.  Secondly, we can take the director
frame $\{ \vec{d}_1, \vec{d}_2, \vec{d}_3 \}$ to be orthonormal, such
that $\vec{d}_1$ is the unit tangent vector $\vec{r}'$.  We will not
take the effect of gravity into account.

In addition, we now assume that the centerline is a planar curve
moving in the horizontal $(x, y)$-plane, i.e. $\vec{r}(t, s)$ can be
written as $(x(t, s), y(t, s), 0)$.  We introduce the \emph{slope}
$\varphi(t,s)$ of the centerline as $(\cos \varphi, \sin \varphi) =
(x'(t,s), y'(t,s))$.  Furthermore, we define the angle $\theta(t, s)$,
referred to as the \emph{torsion} of the rod, as the angle subtended
between $\vec{e}_z$ and $\vec{d}_3$.  The director frame is completely
determined once we know the slope $\varphi(s, t)$ and the torsion
$\theta(s, t)$.

The specific constraints imposed on our rod model therefore allow us
to eliminate the director frame in favour of the slope $\varphi$ and
the torsion $\theta$.  Furthermore, as we shall see, the slope
$\varphi$ is related to the curvature of the centerline.  Note that,
in formulating the dynamics, we still have to impose the
inextensibility condition $(x')^2 + (y')^2 = 1$.

\begin{remark}
  Note that $\theta$ has nothing to do with the usual geometric
  concept of torsion of a curve in $\bfR^3$, and neither is $\theta$
  related to the concept of shear in (for example) the theory of the
  Timoshenko beam.
\end{remark}

\subsection{The dynamics}

As the director frame is orthonormal, there exists a vector $\vec{u}$,
defined by $\vec{d}'_i = \vec{u} \times \vec{d}_i$, called the
\emph{strain} or \emph{Darboux vector}.  With the conventions from the
previous section, $\vec{u}$ takes the following form:
\[
  \vec{u} = \theta' \vec{d}_1 + \varphi' \vec{e}_z.
\]
($\vec{u}$ can be thought of as an ``angular momentum'' vector, but
with time-derivatives replaced by derivatives with respect to
arclength.) 

The dynamics of our rod model can be derived from a variational
principle.  The kinetic energy is given by 
\[
  T = \frac{1}{2} \int_0^\ell\left( \rho(s) ( \dot{x}^2 + \dot{y}^2) + \alpha
  \dot{\theta}^2 \right)\, \d s, 
\]
where $\alpha$ is an appropriately chosen constant.  Here, the mass
density is denoted by $\rho$, and will be assumed constant from now
on.

For a hyperelastic rod, the potential energy is of the form
\[
  V = \int_0^\ell W(u_1, u_2, u_3) \d s, 
\]
where $W(u_1, u_2, u_3)$ is called the \emph{stored energy density},
and the $u_i$ are the components of $\vec{u}$ relative to the director
frame: $u_i = \vec{u} \cdot \vec{d}_i$.  In the simplest case, of
\emph{linear elasticity}, $W$ is a quadratic function of the strains:
\begin{equation} \label{SEfunc}
  W(u_1, u_2, u_3) = \frac{1}{2} \left( K_1 u_1^2 + K_2 u_2^2 + K_3
    u_3^2 \right).
\end{equation}
We will not dwell on the physical interpretation of the constants
$K_i$ any further (in this case, they are related to the moments of
inertia of the laminae).  If the rod is transversely isotropic,
i.e. if the laminae are invariant under rotations around $\vec{d}_1$,
we may take $K_2 = K_3$.  The potential energy then becomes
\begin{equation} \label{potential}
V = \frac{1}{2}\int_0^\ell \left( \beta (\theta')^2 + K \kappa^2 \right)
\d s,
\end{equation}
where $\kappa$ is the curvature of the centerline, i.e. $\kappa^2 =
(\varphi')^2 = (x'')^2 + (y'')^2$, and where we have put $\beta
:= K_1$ and $K := K_2$.  Models with a similar potential energy abound
throughout the literature and are generally referred to as the Euler
elastica.  For more information, see \cite{kirchhoff} and the
references therein.

\subsection{The second-order model}

Having eliminated the derivative of the slope $\varphi$ from the
stored energy density, we end up with a model in which the fields are
the coordinates of the centerline $(x(t, s), y(t, s))$ and the torsion
angle $\theta(t, s)$.  This model fits into the framework developed in
section~\ref{sec:2ndorder}; the base space $X$ is $\bfR^2$, with
coordinates $(t, s)$ and the total space $Y$ is $X \times \bfR^2
\times \bfS^1$, with fibre coordinates $(x, y, \theta)$.

The total Lagrangian now consists of the density of kinetic energy
minus that of potential energy, as well as an additional term
enforcing the constraint of inextensibility, and can be written as
\begin{equation} \label{roddensity} L = \frac{\rho}{2} (\dot{x}^2 +
  \dot{y}^2) + \frac{\alpha}{2} \dot{\theta}^2 - \frac{1}{2}\left(
    \beta (\theta')^2 + K \kappa^2 \right) - \frac{1}{2}p \left(
    (x')^2 + (y')^2 - 1 \right),
\end{equation}
where $p$ is a Lagrange multiplier associated to the constraint of
inextensibility.  The field equations associated to this Lagrangian
take the following form:
\begin{equation} \label{freeFE}
\left\{
  \begin{array}{rcl}
    \rho \ddot{x} + K x'''' & = & \frac{\partial}{\partial s} (p x') \\
    \rho \ddot{y} + K y'''' & = & \frac{\partial}{\partial s} (p y') \\
    \alpha \ddot{\theta} - \beta \theta'' & = & 0, 
   \end{array}
\right.
\end{equation}
to be supplemented with the inextensibility constraint 
\begin{equation} \label{constrinext}
  (x')^2 + (y')^2 = 1,
\end{equation}
which allows to determine the multiplier $p$.  Note in passing that
the dynamics of the centerline and the torsion angle $\theta$ are
completely uncoupled.  This will change once we add nonholonomic
constraints.

\subsection{Field equations and symmetries}

We recall the expression (\ref{2ndcartan}) for the second-order Cartan
form.  If a Lie group $G$ is acting on $Y$ by bundle automorphisms,
and on $J^3\pi$ by prolonged bundle automorphisms, there is a
Lagrangian momentum map $J^L_\xi = \xi_{J^3\pi} \contraction
\Omega_L$, as described in section~\ref{sec:covft}.  We now turn to a
brief overview of the symmetries associated to the rod model
introduced in the previous section.  For an overview of symmetries in
the general theory of Cosserat rods, see \cite{Dichmann}.

\paragraph{Translations in time}

The Lie group $\bfR$ acts on $X$ by translations in time:
$\Phi_\epsilon: (s, t) \mapsto (s, t + \epsilon)$.  The Lagrangian is
invariant and the pullback to $X$ (by a solution $j^3\phi$ of the
field equations) of the momentum map associated to the infinitesimal
generator $\frac{\partial}{\partial t}$ is given by
\begin{eqnarray} \label{ConsLawEnergy}
\fl  (j^3\phi)^\ast J^L_1 =
  \left[  (p x' - K x''') \dot{x} + (py' - Ky''') \dot{y}
    + \beta \theta' \dot{\theta} + K(x'' \dot{x}' + y'' \dot{y}')
  \right] \d t \\
\lo+ \big[ \underbrace{
      \frac{\rho}{2}(\dot{x}^2 + \dot{y}^2) +
      \frac{\alpha}{2} \dot{\theta}^2 + \frac{K}{2}((x'')^2 + (y'')^2) +
      \frac{\beta}{2} (\theta')^2 + \frac{p}{2}((x')^2 + (y')^2 -
      1)}_{\calE}\big] \d s \nonumber,
\end{eqnarray} 
where we have introduced the \emph{energy density} $\calE$.  By taking
the exterior derivative of (\ref{ConsLawEnergy}) and integrating the
conservation law $\d[(j^3\phi)^\ast J^L_1] = 0$  
over $[0, \ell] \times [t_0, t_1] \subset \bfR^2$, we obtain
\begin{equation} \label{totalenergy} \fl E(t_1) - E(t_0) =
  \int_{t_0}^{t_1} \left[ (p x' - K x''') \dot{x} + (py' - Ky''')
    \dot{y} + \beta \theta' \dot{\theta} + K(x'' \dot{x}' + y''
    \dot{y}') \right]^\ell_0 \d t,
\end{equation}
where $E(t) = \int_0^\ell \calE \d s$ is the total energy, which is
conserved if suitable boundary conditions are imposed.  This is the
case, for instance, for periodic boundary conditions or when both ends
of the rod can move freely, i.e. when
\[
p x' - K x''' = p y' - K y''' = 0 \qqand x'' = y'' =
\theta' = 0 \quad \mbox{at $s = 0, \ell$}.
\]

\paragraph{Spatial translations}

Consider the Abelian group $\bfR^2$ acting on the total space $Y$ by
translation, i.e. for each $(a, b) \in \bfR^2$ we consider the map
$\Phi_{(a, b)} : (s, t; x, y, \theta) \mapsto (s, t; x + a, y + b,
\theta)$.  The Lagrangian density is invariant under this action and
the associated momentum map is
\[
(j^3\phi)^\ast J^L_{(v_1, v_2)} = - \rho(v_1\dot{x} + v_2 \dot{y}) \d s -
    (v_1p x' - v_1Kx''' + v_2p y' - v_2K y''') \d t 
\]
for all $(v_1, v_2) \in \bfR^2$.  Again, under suitable boundary
conditions, $J^L_{(v_1, v_2)}$ gives rise to a conserved quantity,
namely the total linear momentum of the rod.

Similarly, $\bfS^1$ acts on $Y$ by translations in $\theta$, with
infinitesimal generator of the form $\pdd{\theta}$, and the
corresponding momentum map is
\[
  (j^3\phi)^\ast J^L_1 = -\beta \theta' \d t - \alpha \dot{\theta} \d s.
\]
The ensuing conservation law is given by $\alpha \ddot{\theta} = \beta
\theta''$ and, hence, is just the equation of motion for $\theta$.

\paragraph{Spatial rotations}

Finally, we note that the rotation group $SO(2)$ acts on $Y$ by
rotations in the $(x, y)$-plane.  The infinitesimal generator
corresponding to $1 \in \gothso(2) \cong \bfR$ is given
by $y \frac{\partial}{\partial x} - x \frac{\partial}{\partial y}$;
its prolongation to $J^3 \pi$ is
\[
\xi_{J^3\pi} = y \frac{\partial}{\partial x} - x
\frac{\partial}{\partial y} + \dot{y} \frac{\partial}{\partial
  \dot{x}} - \dot{x}\frac{\partial}{\partial \dot{y}} + y'
\frac{\partial}{\partial x'} - x' \frac{\partial}{\partial y'} +
\cdots,
\]
where the dots represent terms involving higher-order derivatives.  As
$\Theta_L$ is semibasic with respect to $\pi_{3, 1}$, these terms make
no contribution to the momentum map.  The momentum map is given by 
\[
\fl (j^3\phi)^\ast J^L_1 = \left[ -x(-py' + Ky''') + y(-px' + Kx''') -
  K(x''y' + y''x') \right] \d t + \rho( x \dot{y} - y \dot{x}) \d s,
\]
leading to the conservation of total angular momentum.  Note that the
angular momentum does not involve $\theta$, in contrast to the
corresponding expression in more general treatments of Cosserat
media.  This is a consequence of the fact that we defined the action
of $SO(2)$ on $Y$ to act trivially on the $\theta$ part.

\subsection{A nonholonomic model} \label{sec:NH}

Consider again a Cosserat rod as in the previous section.  The
constraint that we are now about to introduce is a generalization of
the familiar concept of rolling without sliding in mechanics: we
assume that the rod is placed on a horizontal plane, which we take to
be perfectly rough, so that each of the laminae rolls without sliding.

However, as the Cosserat rod is also supposed to be incompressible,
one must take care that the additional constraints do not become too
restrictive.\footnote{This was pointed out to me by W. Tulczyjew and
  D. Zenkov.}  Indeed, a simple argument shows that the model of an
incompressible rod which rolls without sliding, and which cannot move
transversally, can only move like a rigid body.

There are two immediate solutions: either one relaxes the
incompressibility constraint, or one allows the rod to move laterally
as well.  Either solution introduces a lot of mathematical tedium
which greatly obscures the physical background of the system.  For
this paper, we will therefore consider a simplified model containing
aspects of both models.

In particular, we will \emph{assume} that the motion of the
nonholonomic rod is such that the incompressibility constraint is
satisfied approximately throughout the motion; this is equivalent to
the following assumption:
\begin{equation}
  \sqrt{(x')^2 + (y')^2} \cong 1.
\end{equation}
By neglecting the incompressibility constraint in the Lagrangian, a
simplified model is then obtained.  Of course, this new model is a
mathematical simplification of the true physics.  However, numerical
simulations show that $\sqrt{(x')^2 + (y')^2}$ is bounded throughout
the motion, and it's therefore reasonable that the dynamics of this
model is close to the true dynamics.  One could think of the
mathematical model as describing a Cosserat rod whose constitutive
equation is specified on mathematical grounds, rather than derived
from first principles.

The constraints of rolling without sliding are given by (see
\cite{bloch, cortes}):
\begin{equation} \label{constraintsphi}
  \dot{x} + R \dot{\theta} \sin \varphi = 0 \qqand \dot{y} - R
  \dot{\theta} \cos\varphi = 0,
\end{equation}
where $R$ is the radius of the laminae.  By eliminating the slope
$\varphi$ we then obtain
\begin{equation} \label{constraints}
  \dot{x} + R \dot{\theta} y' = 0 \qqand \dot{y} - R \dot{\theta} x' =
  0. 
\end{equation}
Incidentally, the passage from (\ref{constraintsphi}) to
(\ref{constraints}) again illustrates why derivatives with respect to
time play a fundamentally different role as opposed to the other
derivatives.

The Lagrangian density of the nonholonomic rod is still given by
(\ref{roddensity}); we recall that it is of second order, as the
stored energy function (\ref{SEfunc}) is of grade two.  The constraint
on the other hand is of first order.  By demanding that the action be
stationary under variations compatible with the given constraint (a
similar approach to section~\ref{sec:nonholft}), we obtain the
following field equations:

\begin{definition}
  A section $\phi$ of $\pi$ is a solution of the nonholonomic problem
  if and only if $\im j^1\phi \subset \calC$, and, along $\calC$, 
  \begin{equation} \label{nonhFE}
  (j^3 \phi)^\ast (j^3 V \contraction \Omega_L) = 0
  \end{equation}
  for all $\pi$-vertical vector fields $V$ on $Y$ such that
  $(j^1\phi)^\ast(j^1 V \contraction \alpha) = 0$ for all $\alpha \in
  F$.
\end{definition}

The left-hand side of (\ref{nonhFE}) is just the Euler-Lagrange
equation (\ref{2ndEL}) for a second-order Lagrangian.  As the
constraint is first order, it can be treated exactly as in
section~\ref{sec:nonholft}. In coordinates, the nonholonomic field
equations hence are given by
\[
\left[\frac{\partial L}{\partial y^a} - \frac{\d}{\d x^\mu} \left(
    \frac{\partial L}{\partial y^a_\mu} \right) + \frac{\d^2}{\d x^\mu
    \d x^\nu} \left( \frac{\partial L}{\partial y^a_{\mu\nu}}
  \right)\right](j^4\phi) = \lambda_\alpha \frac{\partial
  \varphi^\alpha}{\partial y^a_0}(j^1\phi).
\]

By substituting the Lagrangian (\ref{roddensity}), without the
inextensibility constraint, and the constraints (\ref{constraints})
into the Euler-Lagrange equations, we obtain the following set of
nonholonomic field equations:
\begin{equation} \label{NHFE}
\left\{
  \begin{array}{rcl}
    \rho \ddot{x} + K x'''' & = & \lambda\\
    \rho \ddot{y} + K y'''' & = & \mu \\
    \alpha \ddot{\theta} - \beta \theta'' & = & R(\lambda y' + \mu x')
   \end{array}
\right.
\end{equation}
where $\lambda$ and $\mu$ are Lagrange multipliers associated with
the nonholonomic constraints.  These equations are to be supplemented by the
constraint equations (\ref{constraints}). 

In the familiar case of the rolling disc, it is well known that energy
is conserved.  There is a similar conservation law for the
nonholonomic rod.

\begin{prop} \label{prop:NHenergy} The total energy
  (\ref{totalenergy}) is conserved for each solution of the
  nonholonomic field equations (\ref{NHFE}) and constraints
  (\ref{constraints}).  A fortiori, the solutions of the nonholonomic
  field equations satisfy the local conservation law $\d[
  (j^3\phi)^\ast J^L_1] = 0$, where $J^L_1$ is the momentum map
  associated to time translation introduced in (\ref{ConsLawEnergy}).
\end{prop}
\begin{mproof}
This follows immediately from proposition~\ref{NHnoether} in the
appendix, and the fact that $\frac{\partial}{\partial t}$ (or rather
its prolongation to $J^1\pi$) annihilates $\bar{F}$ along the constraint
manifold.  Indeed, the bundle of $(n + 1)$-forms $\bar{F}$ is
generated by $\Phi^1$ and $\Phi^2$, defined as follows:
\begin{eqnarray*}
\Phi^1 = (\d x - \dot{x}\d t) \wedge \d s + Ry'(\d\theta -
\dot{\theta} \d t) \wedge \d s; \\
 \Phi^2 = (\d y - \dot{y}\d t)
\wedge \d s - Rx'(\d\theta - \dot{\theta} \d t) \wedge \d s.
\end{eqnarray*}
Therefore, we have that
\[
\left( \frac{\partial}{\partial t} \right)_{J^1\pi} \contraction
\Phi^1 = - (\dot{x} + R\dot{\theta}y') \d s,
\]  
which vanishes when restricted to $\calC$.  A similar argument shows
that the contraction of $\frac{\partial}{\partial t}$ with $\Phi^2$
vanishes.  Hence, proposition~\ref{NHnoether} can be applied; the
associated momentum map is just (\ref{ConsLawEnergy}).
\end{mproof}

\section{Discrete nonholonomic field theories} \label{sec:numerical}

In this section we present an extension to the case of field theories
of the discrete d'Alembert principle described in
\cite{nhdiscreet}. We also study an elementary numerical integration
scheme aimed at integrating the field equations (\ref{NHFE}).  

As in the previous sections, we will consider the trivial bundle $\pi$
with base space $X = \bfR \times M$ (where $M = [0, \ell]$), and total
space $Y$ the product $X \times S$, where $S = \bfR^2 \times
\bfS^1$. Our discretization scheme is the most straightforward
possible: The base space $X$ will be discretized by means of the
uniform mesh $\bfZ \times \bfZ$, and the total space $Y$ by replacing
it with $X \times \bfR^2 \times \bfR$.

\subsection{Discrete Lagrangian field theories}

We begin by giving an overview of discrete Lagrangian field theories,
inspired by \cite{Kouranbaeva, MPS98}.  In order to discretize the
second-order jet bundle, we need to approximate the derivatives of the
field (of first and second order).  This we do by means of central
differences with spatial step $k$ and time step $h$:
\begin{equation} \label{FDeta} 
\fl  \dot{\eta} \approx \frac{\eta_{n+1, i} - \eta_{n, i}}{h}, \quad
  \eta' \approx \frac{\eta_{n, i+1} - \eta_{n, i-1}}{2k}, \qqand
  \eta'' \approx \frac{\eta_{n, i+1} - 2\eta_{n, i} + \eta_{n, i-1}}{k^2},
\end{equation}
where $\eta$ stands for either $x$ or $y$.  Other derivatives will not
be needed.  For $\theta$, we use
\begin{equation} \label{FDtheta}
  \dot{\theta} \approx \frac{\theta_{n+1, i} - \theta_{n, i}}{h}
  \qqand \theta' \approx \frac{\theta_{n, i+1} - \theta_{n, i}}{k}. 
\end{equation}

Let $\calM$ be the uniform mesh in $X = \bfR^2$ whose elements are
points with integer coordinates; i.e. $\calM = \bfZ \times \bfZ$.  The
elements of $\calM$ are denoted as $(n, i)$, where the first component
refers to time, and the second to the spatial coordinate.  We define a
\emph{$9$-cell} centered at $(n, i) \in \calM$, denoted by $[x]_{(n,
  i)}$, to be a nine-tuple of the form
\begin{eqnarray} 
  [x]_{(n,i)} & := &\big( (n-1, i-1), (n-1, i), (n-1, i); 
    (n, i-1), \label{cell} \\
    & & (n, i), (n, i+1); (n+1, i-1), (n+1, i), (n+1, i+1) \big)
    \nonumber 
\end{eqnarray}
It is clear from the finite difference approximations that a generic
second-order jet $j^2_x\phi$ can be approximated by specifying the
values of $\phi$ at the nine points of a cell.

However, in the case of the nonholonomic rod, the Lagrangian depends
only on the derivative coordinates whose finite difference
approximations were given in (\ref{FDeta}) and (\ref{FDtheta}).
Therefore, we can simplify our exposition by defining a
\emph{$6$-cell} at $(n, i)$ to be the six-tuple
\begin{equation}
 \fl [x]_{(n,i)} := \label{6cell} 
\big((n, i-1), (n, i), (n, i+1); 
    (n+1, i-1), (n+1, i), (n+1, i+1) \big).
\end{equation}
We will refer to $6$-cells simply as \emph{cells}.  Let us denote the
set of all cells by $\bfX^6 := \{ [x]_{(n, i)} : (n, i) \in \calM \}$.
We now define the \emph{discrete $2$nd order jet bundle} to be $J^2_d
\pi := \bfX^6 \times S^6$ (see \cite{Kouranbaeva, MPS98, discreet}). A
\emph{discrete section} of $\pi$ (also referred to as a \emph{discrete
  field}) is a map $\phi: \calM \rightarrow S$.  Its \emph{second jet
  extension} is the map $j^2\phi: \bfX^6 \rightarrow J^2_d\pi$ defined as
\[
  j^2\phi([x]_{(n,i)}) := ([x]_{(n, i)}; \phi(x_1), \ldots \phi(x_6)),
\]
where $x_1, \ldots, x_6$ are the vertices that make up $[x]_{(n, i)}$
(ordered as in (\ref{6cell})).  Given a vector field $W$ on $Y$, we
define its second jet extension to be the vector field $j^2 W$ on
$J^2_d$ given by
\[
  j^2W([x]; f_1, \ldots, f_6) = (W(x_1, f_1), W(x_2, f_2), \ldots, 
    W(x_6, f_6)).
\]

Let us now assume that a \emph{discrete Lagrangian} $L_d: J^2_d\pi
\rightarrow \bfR$ is given.  The \emph{action sum} $S_d$ is then defined
as
\begin{equation} \label{actionsum}
  S_d(\phi) = \sum_{[x]} L_d( j^2\phi([x]) ).
\end{equation}
Given a vertical vector field $V$ on $Y$ and a discrete field $\phi$,
we obtain a one-parameter family $\phi_\epsilon$ by composing $\phi$
with the flow $\Phi$ of $V$: 
\begin{equation} \label{discvar}
\phi_\epsilon([x]) = \left([x]; \Phi_\epsilon(\phi([x])_1), \ldots,
\Phi_\epsilon(\phi([x])_6)\right).
\end{equation}

The \emph{variational principle} now consists of seeking discrete
fields $\phi$ that extremize the discrete action sum.  The fact that
$\phi$ is an extremum of $S$ under variations of the form
(\ref{discvar}) is expressed by
\begin{eqnarray}
\fl \sum_{(n,i)\in\calM} \Big< X(\phi_{(n, i)}), D_1 L(j^2 \phi([x]_{(n, i+1)}))
  + D_2 L(j^2 \phi([x]_{(n, i)})) + D_3 L(j^2 \phi([x]_{(n, i-1)})) 
  \label{variations} \\
\lo+  D_4 L(j^2 \phi([x]_{(n-1, i+1)})) + D_5
  L(j^2 \phi([x]_{(n-1, i)})) + D_6 L(j^2 \phi([x]_{(n-1, i-1)})) \Big> =
  0. \nonumber
\end{eqnarray}
As the variation $X$ is completely arbitrary, we obtain the following
set of \emph{discrete Euler-Lagrange field equations}:
\begin{eqnarray}
  D_1 L(j^2 \phi([x]_{(n, i+1)})) + D_2 L(j^2 \phi([x]_{(n, i)})) + D_3
  L(j^2 \phi([x]_{(n, i-1)})) + \label{discEL} \\
  D_4 L(j^2 \phi([x]_{(n-1, i+1)})) + D_5
  L(j^2 \phi([x]_{(n-1, i)})) + D_6 L(j^2 \phi([x]_{(n-1, i-1)})) =
  0. \nonumber
\end{eqnarray}

for all $(n, i)$.  Here, we have denoted the values of the field
$\phi$ at the points $(n, i)$ as $\phi_{n, i}$.

\subsection{The discrete d'Alembert principle}

Our discrete d'Alembert principle is nothing more than a suitable
field-theoretic extension of the discrete Lagrange-d'Alembert
principle described in \cite{nhdiscreet}.  Just as in that paper, in
addition to the discrete Lagrangian $L_d$, two additional ingredients
are needed: a discrete constraint manifold $\calC_d \subset J^1_d \pi$
and a bundle of constraint forces $F_d$ on $J^2_d\pi$.  However, as
our constraints (in particular (\ref{constraints})) are not linear in
the derivatives, as opposed to the case in \cite{nhdiscreet}, our
analysis will be more involved.

The discrete constraint manifold $\calC_d \hookrightarrow J^1_d\pi$
will usually be constructed from the continuous constraint manifold
$\calC$ by subjecting it to the same discretization as used for the
discretization of the Lagrangian (i.e. (\ref{FDeta}) and
(\ref{FDtheta})).  To construct the discrete counterpart $F_d$ of the
bundle of discrete constraint forces, somewhat more work is needed.

\begin{remark}
  For the discretization of the constraint manifold, it would appear
  that we need a discrete version of the first-order jet bundle as
  well.  A similar procedure as for the discretization of the
  second-order jet bundle (using the same finite differences as in
  (\ref{FDeta}) shows that a discrete $1$-jet depends on the values of
  the field at the same four points of a cell as a discrete $2$-jet:
  the difference between $J^1_d\pi$ and $J^2_d\pi$ lies in the way in
  which the values of the field at these points are combined.
  Therefore, we can regard the discrete version of $\calC$, to be
  defined below, as a subset of $J^2_d\pi$.
\end{remark}

\subsubsection{The bundle of discrete constraint forces}

In this section, we will construct $F_d$ by following a discrete
version of the procedure used in section~\ref{sec:nonholft}.  Just as
in the continuous case, it is here that the difference between spatial
and time derivatives will become fundamental.  Indeed, we will
discretize with respect to space first, and (initially) not with
respect to time.  It should be noted that the construction outlined in
this paragraph is not entirely rigorous but depends strongly on
coordinate expressions.  Presumably, one would need a sort of
\emph{discrete Cauchy analysis} in order to solidify these arguments.
For now, we will just accept that this reasoning provides us with the
correct form of the constraint forces.

For the sake of convenience, we suppose that $\calC$ is given by the
vanishing of $k$ independent functions $\varphi^\alpha$ on $J^1\pi$.
By applying the \emph{spatial} discretizations in (\ref{FDeta}) and
(\ref{FDtheta}) to $\varphi^\alpha$, we obtain $k$ functions, denoted
as $\varphi^\alpha_{1/2}$, on $J^2_d \times TS$.   We define the
\emph{semi-discretized constraint submanifold} $\calC_{1/2}$ to be the
zero level set of the functions $\varphi^\alpha_{1/2}$. 

Consider now the forms 
\[
\Phi^\alpha_{1/2} := J^\ast (\d \varphi^\alpha_{1/2})
\] 
(where $J$ is the vertical endomorphism on $TS$); they are the
semi-discrete counterparts of the forms $\Phi^\alpha$ defined in
(\ref{constrforms}).  The forms $\Phi^\alpha_{1/2}$ are semi-basic.  By
discretizing the time derivatives, however, we obtain a set of basic
forms on $J^2_d\pi$, which we also denote by $\Phi^\alpha_{1/2}$.  An
example will make this clearer.

\begin{example}  
  Consider, for instance, the constraint manifold $\calC
  \hookrightarrow J^2\pi$ defined as the zero level set of the
  function $\varphi = A_{ab}\,y^a_0 y^b_1 + B_b (y^b_1)^2$, where
  $A_{ab}$ and $B_b$ are constants.  By applying (\ref{FDeta}) and
  (\ref{FDtheta}), it follows that $\calC_d$ is given as the zero
  level set of the function
  \[ 
  \varphi_d([y]) := A_{ab} \frac{y^a_{n+1, i} - y^a_{n, i}}{h}
  \frac{y^b_{n, i+1} - y^b_{n, i-1}}{2k} + B_b \left( \frac{y^b_{n, i+1}
      - y^b_{n, i-1}}{2k} \right)^2 
  \]
  for $[y] \in J^1_d \pi$,
  and $\calC_{1/2}$ as the zero level set of the function
  \[ 
  \varphi_{1/2}([y], v) := A_{ab}\,\dot{v}^a\, \frac{y^b_{n, i+1} -
    y^b_{ n, i-1}}{2k} + B_b \left( \frac{y^b_{n, i+1} - y^b_{n,
        i-1}}{2k} \right)^2 
  \]
  for $[y] \in J^1_d \pi$ and $v \in TS$.  The bundle $F_d$ is then
  generated by the one-form $\Phi_{1/2} := S^\ast( \d \varphi_{1/2})$, or
  explicitly,
  \[
    \Phi = A_{ab} \frac{y^b_{n, i+1} - y^b_{n, i-1}}{2k} \d y^a.
  \]
\end{example}

\subsubsection{The discrete nonholonomic field equations}

Assuming that $L_d$, $\calC_d$ and $F_d$ are given (their construction
will be treated in more detail in the next section), the derivation of
the discrete nonholonomic field equations is similar to the continuum
derivation: we are looking for a discrete field $\phi$ such that $\im
j^1\phi \subset \calC_d$ and such that $\phi$ is an extremum of
(\ref{actionsum}) for all variations \emph{compatible with the
  constraints}, in the sense that the variation $X$ satisfies, for all
$(n, i)$,
\[
  X(\phi_{(n, i)}) \contraction \Phi^\alpha_{1/2}( j^2\phi( [x]_{(n, i)} ) )
  = 0.
\]
From (\ref{variations}) we then obtain the \emph{discrete nonholonomic
  field equations}:
\begin{eqnarray}
\fl  D_1 L(j^2 \phi([x]_{(n, i+1)})) + D_2 L(j^2 \phi([x]_{(n, i)})) + D_3
  L(j^2 \phi([x]_{(n, i-1)})) + D_4 L(j^2 \phi([x]_{(n-1, i+1)})) \nonumber \\
  \lo+ D_5
  L(j^2 \phi([x]_{(n-1, i)})) + D_6 L(j^2 \phi([x]_{(n-1, i-1)})) =
  \lambda_\alpha \Phi^\alpha_{1/2}(j^2 \phi([x]_{(n, i)})), \label{discNH}
\end{eqnarray}
where the Lagrange multipliers $\lambda_\alpha$ are to be determined
from the requirement that $\im j^2 \phi \subset \calC_d$.

\subsection{An explicit, second-order algorithm} \label{explalg}

In this section, we briefly present some numerical insights into the
nonholonomic field equations of section~\ref{sec:NH}.  Our aim is
twofold: for generic boundary conditions, the nonholonomic field
equations (\ref{NHFE}) cannot be solved analytically and in order to
gain insight into the behaviour of our model, we therefore turn to
numerical methods.  Secondly, in line with the fundamental tenets of
geometric integration, we wish to show that the construction of
practical integration schemes is strongly guided by geometric
principles.

In discretizing our rod model, we effectively replace the continuous
rod by $N$ rigid rolling discs interconnected by some potential (see
\cite{Barth}).  This is again an illustration of the fact that the
constraints are truly nonholonomic.  Our integrator is just a
concatenation of the leapfrog algorithm for the spatial part, and a
nonholonomic mechanical integrator for the integration in time.

As a first attempt at integrating (\ref{NHFE}), we present an
explicit, second-order algorithm.  In the Lagrangian, the derivatives
are approximated by
\[
  \dot{x} \approx \frac{x_{n+1, i} - x_{n, i}}{h} \qqand
  x'' \approx \frac{x_{n, i-1} - 2x_{n, i} + x_{n+1, i}}{k^2}, 
\]
where $h$ is the time step, and $k$ is the space step.  Similar
approximations are used for the derivatives of $y$, and for $\theta$
we use 
\begin{equation} \label{forwarddiff}
  \dot{\theta} \approx \frac{\theta_{n+1, i} - \theta_{n, i}}{h}
  \qqand 
  \theta' \approx \frac{\theta_{n, i+1} - \theta_{n, i}}{k}.
\end{equation}
The discrete Lagrangian density can then be found by substituting
these approximations into the continuum Lagrangian (\ref{roddensity}).
Explicitly, it is given by
\begin{eqnarray}
\fl L_d = \frac{\rho}{2h^2} \left( (x_{n+1, i} - x_{n, i})^2 + (y_{n+1, i} -
    y_{n, i})^2\right) + \frac{\alpha}{2h^2} (\theta_{n+1, i} - \theta_{n,
    i})^2 - \frac{\beta}{2k^2} (\theta_{n, i+1} - \theta_{n,
    i})^2 \nonumber \\
  \lo- \frac{K}{2k^4} (x_{n, i-1} - 2x_{n,i} + x_{n, i+1})^2 -
  \frac{K}{2k^4} 
  (y_{n, i-1} - 2y_{n, i} + y_{n, i+1})^2. \label{discretelagrangian} 
\end{eqnarray}
Note that $L_d$ only depends on four of the six points in each cell
(see (\ref{6cell})).  The discrete constraint manifold $\calC_d$ is
found by discretizing the constraint equations (\ref{constraints}).
In order to obtain a second-order accurate approximation, we use
central differences: 
\[
  x' \approx \frac{x_{n, i+1} - x_{n, i-1}}{2k},
\]
(and similar for $y', \dot{x}, \dot{y}, \dot{\theta}$) and hence we
obtain that $\calC_d$ is given by
\begin{equation} \label{discconstr1}
  x_{n + 1, i} - x_{n-1, i} + \frac{R}{2k}(\theta_{n+1, i} - \theta_{n-1,
    i})(y_{n, i+1} - y_{n, i-1}) = 0,
\end{equation}
and
\begin{equation} \label{discconstr2}
  y_{n + 1, i} - y_{n-1, i} - \frac{R}{2k}(\theta_{n+1, i} - \theta_{n-1,
    i})(x_{n, i+1} - x_{n, i-1}) = 0,
\end{equation}
for all $(n, i)$.  The semi-discrete constraint manifold
$\calC_{1/2}$, on the other hand, is given by 
\[ 
\dot{x}_{n, i} + \frac{R}{2k} \dot{\theta}_{n, i}(y_{n, i+1} - y_{n,
  i-1}) = 0, 
\]
and
\[
\dot{y}_{n, i} - \frac{R}{2k} \dot{\theta}_{n,
  i}(x_{n, i+1} - x_{n, i-1}) = 0,
\]
and hence $F_d$ is generated by
\[ 
\Phi^1 = \d x + \frac{R}{2k} (y_{n, i+1} - y_{n, i-1}) \d \theta
\qqand
\Phi^2 = \d y - \frac{R}{2k} (x_{n, i+1} - x_{n, i-1}) \d \theta.
\]

We conclude that the discrete nonholonomic field equations
(\ref{discNH}) are in this case
\begin{equation}
    x_{n+1, i} - 2x_{n, i} + x_{n-1, i} = h^2\lambda_i -
    \frac{h^2K}{k^4} \Delta^4 x_{n, i}
\end{equation}
and
\begin{equation}
    y_{n+1, i} - 2y_{n, i} + y_{n-1, i} = h^2 \mu_i -
    \frac{h^2K}{k^4} \Delta^4 y_{n, i}
\end{equation}
as well as 
\begin{equation*} 
\fl    \alpha( \theta_{n+1, i} - 2\theta_{n, i} + \theta_{n-1, i}) =
      R h^2 \left(\lambda_i \frac{y_{n, i+1} - y_{n, i-1}}{2k} - \mu_i
      \frac{x_{n, i+1} - x_{n, i-1}}{2k}\right) + \frac{\beta
      h^2}{k^2} \Delta^2 \theta_{n, i},
\end{equation*}
where $\Delta^2$ and $\Delta^4$ are the 2nd and 4th order finite
difference operators in the spatial direction, respectively:
\[ 
  \Delta^2 f_{n, i} := f_{n, i+1} - 2f_{n, i} + f_{n, i-1} 
\]
and
\[
\Delta^4 f_{n, i} := f_{n, i+2} - 4f_{n, i+1} + 6f_{n, i} - 4f_{n,
  i-1} + f_{n, i-2}.
\]
In order to determine $\lambda_i$ and $\mu_i$, these equations need to
be supplemented by the discrete constraints (\ref{discconstr1}) and
(\ref{discconstr2}). 

For the purpose of numerical simulation, the following values were
used: $\alpha = 1$, $\beta = 0.8$, $\rho = 1$, $K = 0.7$, $\ell = 4$,
and $R = 1$.  For the spatial discretization, $32$ points were used
(corresponding to $k \approx 0.1290$) and the time step was set to $h
= 1/8k^2$, a fraction of the maximal allowable time step for the
Euler-Bernoulli beam equation (see \cite{Ames}).  The ends of the rod
were left free and the following initial conditions were used:
\[
\vec{r}_0(s) = (s, 0), \quad \theta_0(s) = -\frac{\pi}{2}\cos
\frac{\pi s}{\ell} \qqand \dot{\vec{r}}_0(s) = (0, 0), \quad
\dot{\theta}_0(s) = 0.
\]

\begin{figure}
  \begin{center}
    \includegraphics[scale=0.5]{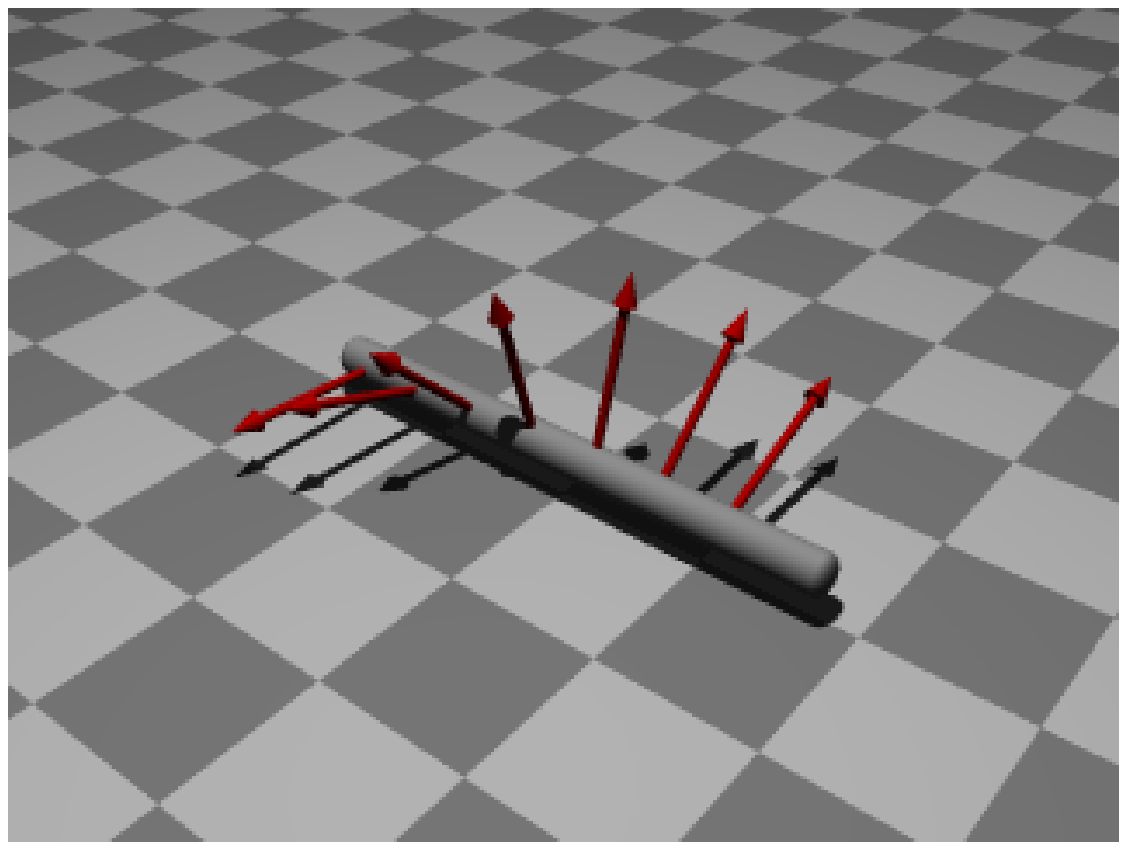} \hspace{0.3in}
    \includegraphics[scale=0.5]{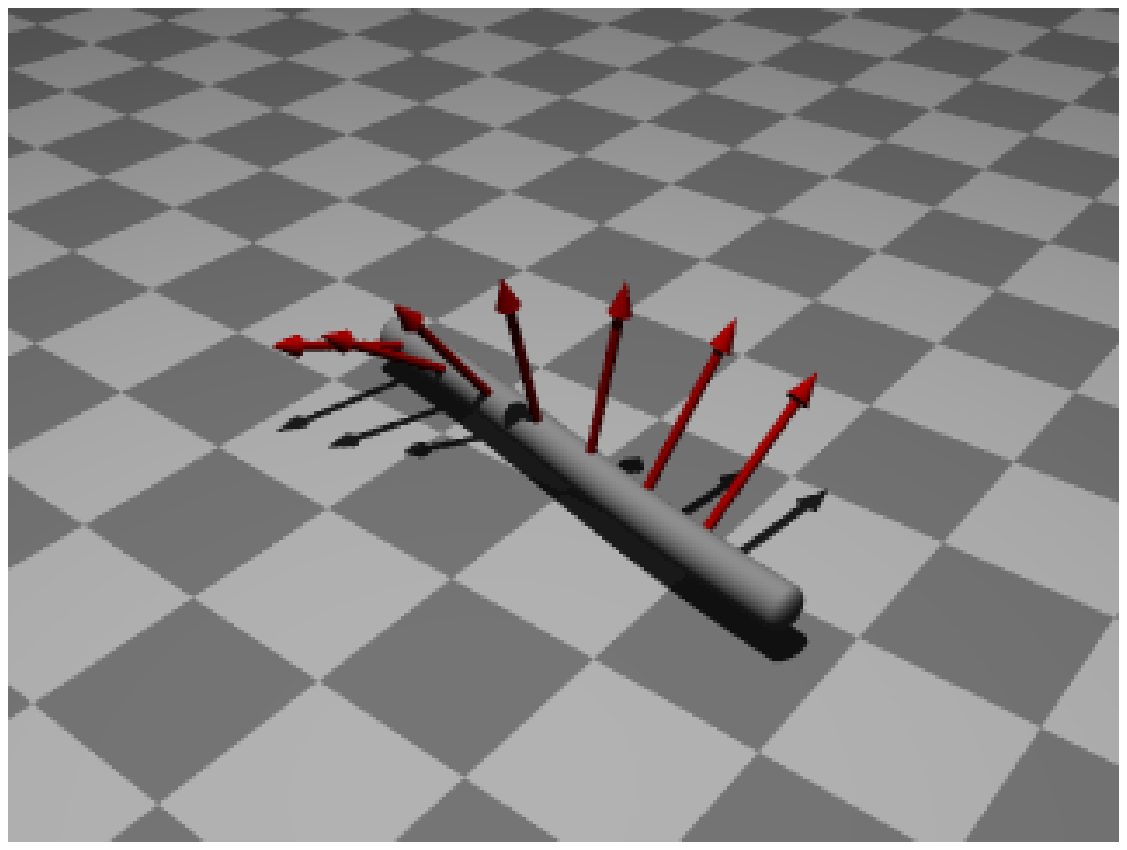} \hspace{0.3in} \\
    \vspace{0.3in}
    \includegraphics[scale=0.5]{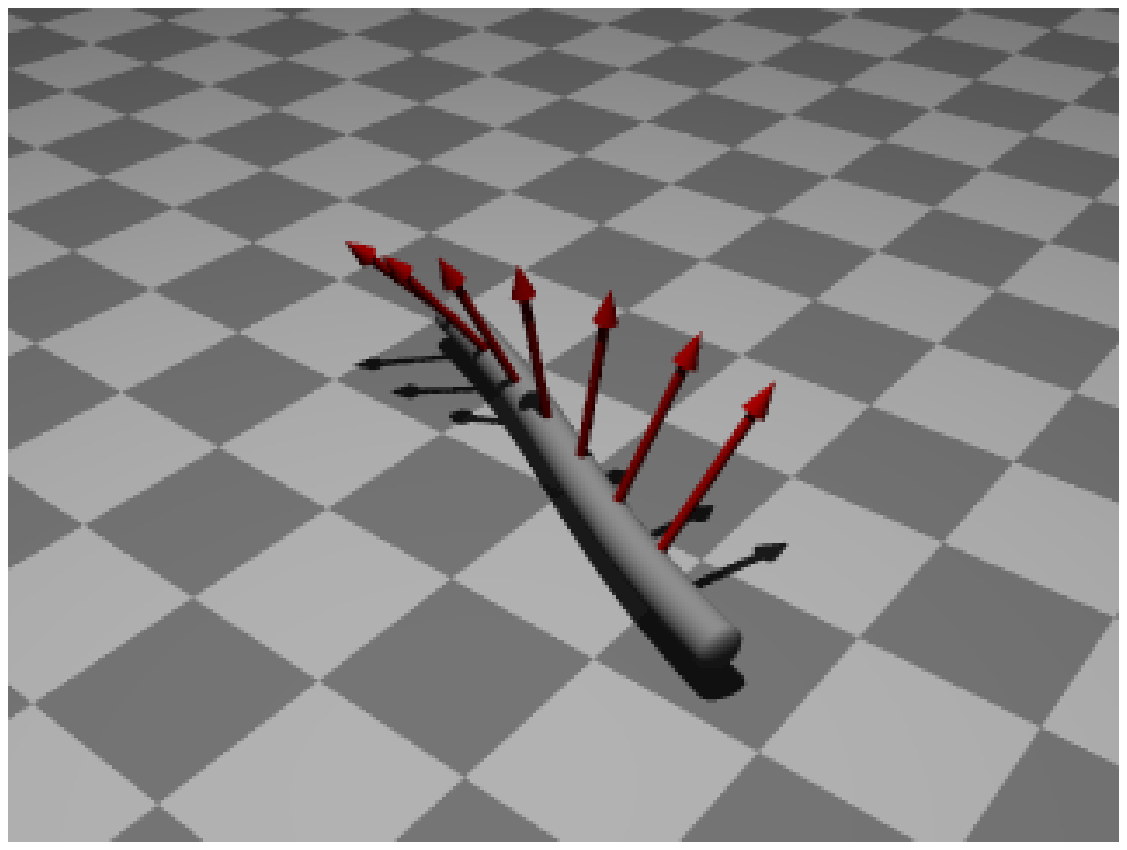} \hspace{0.3in}
    \includegraphics[scale=0.5]{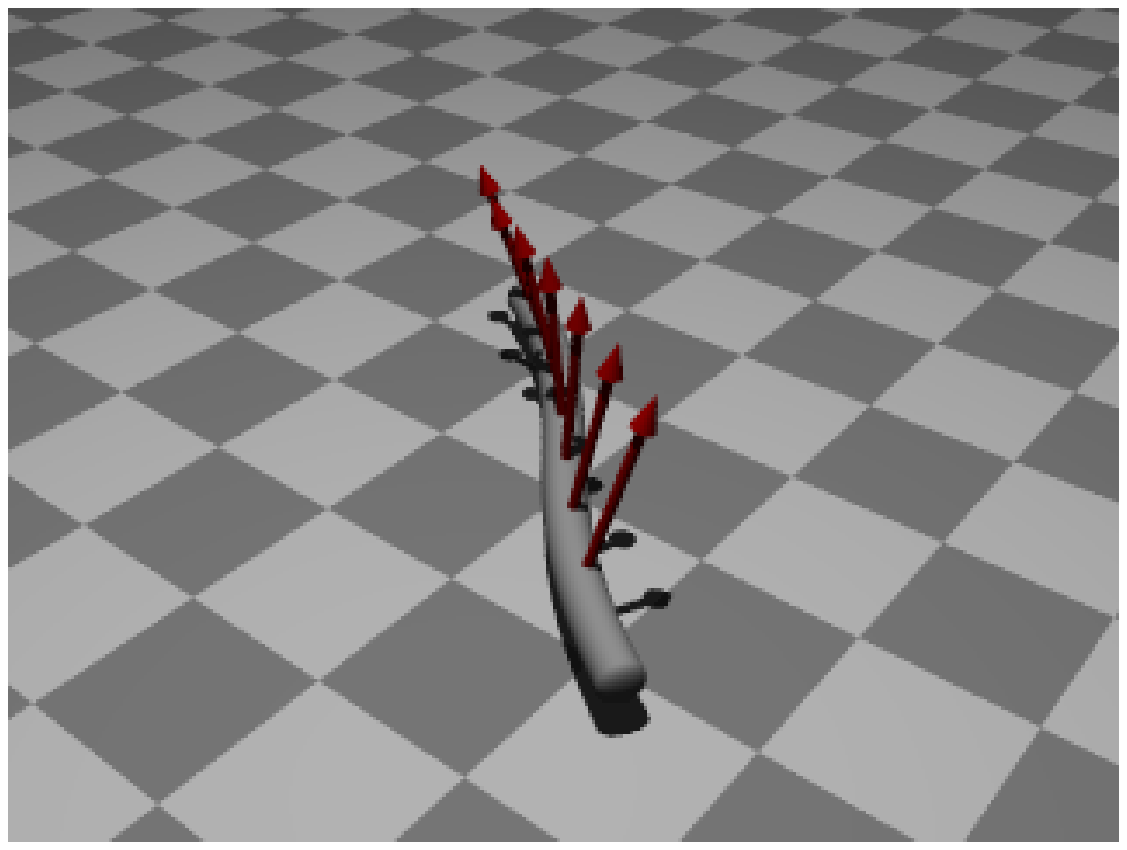} \hspace{0.3in} \\
    \vspace{0.3in}
    \includegraphics[scale=0.5]{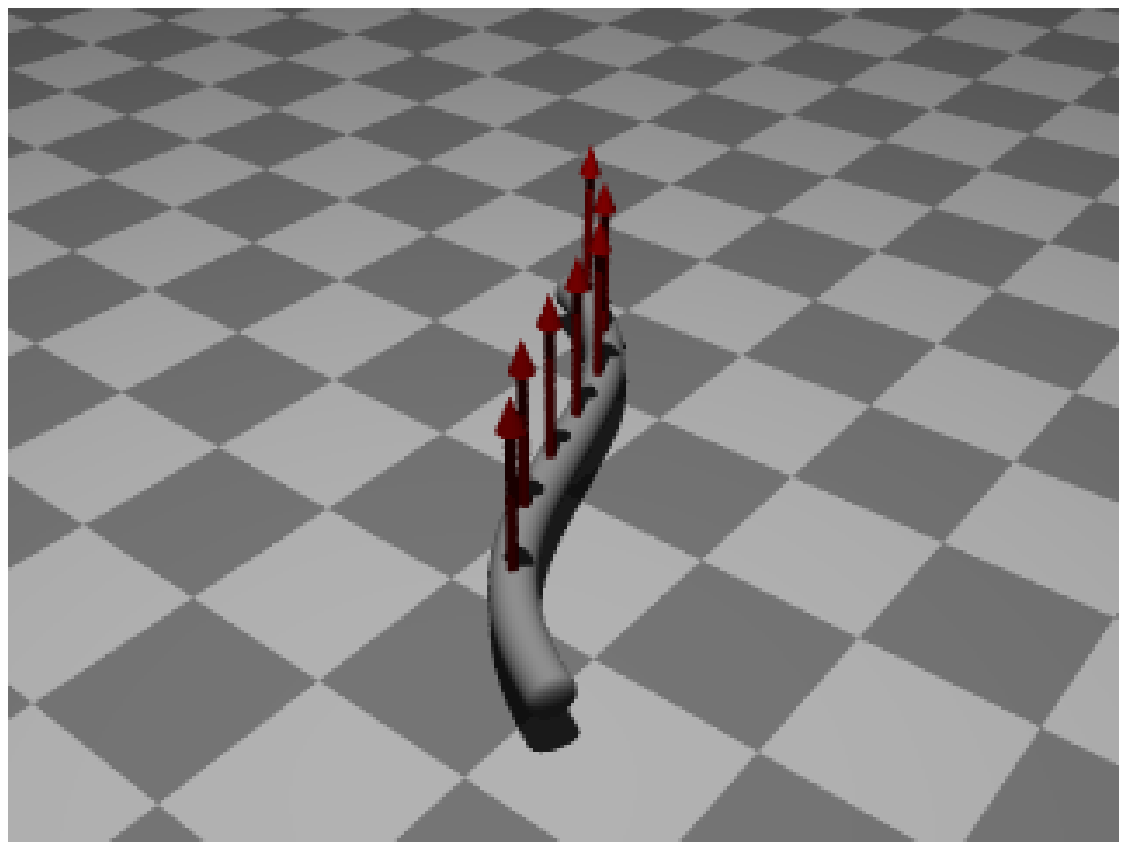}
    \caption{Motion of the rod from $t = 0$ to $t \approx 4.5$.}
    \label{fig:trackrod}
  \end{center}
\end{figure}

An mpeg movie (created with Povray, an open source raytracer)
depicting the motion of the nonholonomic Cosserat rod is available
from the author's web
page\footnote{\url{http://users.ugent.be/$\sim$jvkersch/nonholonomic/}}.
In figure~\ref{fig:trackrod}, an impression of the motion of the rod
is given.  The arrows represent the director field $\vec{d}_3$ and
serve as an indication of the torsion.  The rod starts from an
initially straight, but twisted state and gradually untwists,
meanwhile effecting a rotation.

\begin{figure}
  \begin{center}
    \includegraphics[scale=0.5]{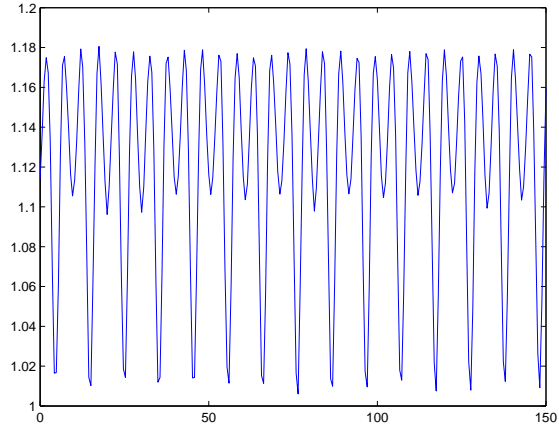}
    \caption{Energy behaviour of the integration algorithm on moderate
    time scales (the interval $[0, 150]$).}
    \label{fig:energy}
  \end{center}
\end{figure}

In figure~\ref{fig:energy}, the energy of the nonholonomic rod is
plotted.  Even though our algorithm is by its very nature \emph{not}
symplectic (or multi-symplectic -- see \cite{Bridges01}), there is
still the similar behaviour of ``almost'' energy conservation.

\section{Conclusions}

It is clear that the study of nonholonomic field theories forms a vast
subject.  This paper gives only a brief survey of a number of
straightforward results, but there are many more things to be
explored.  An acute problem is the lack of an extensive number of
interesting examples; while this of course impedes progress on the
theoretical front, there are nevertheless a number of points worth
investigating, which we now discuss.

In proposition~\ref{prop:NHenergy} we used the fact that the bundle of
reaction forces is annihilated by the generator of time translations
in order to prove conservation of energy.  Even when this is not the
case, experience from mechanics (see \cite{BatesSniatycki, bloch96,
  frans}) as well as from different types of nonholonomic field
theories (see \cite{Vankerschaver05}) seems to suggest that it might
be possible to prove a \emph{nonholonomic momentum equation} instead.

From a numerical point of view, the explicit algorithm of
section~\ref{explalg} is not very accurate.  It is second-order in
space and time and suffers from a restrictive stability condition.
The development of more sophisticated integration schemes that exactly
preserve the nonholonomic constraints would definitely be very
interesting.  Perhaps the most interesting of all, at least in line
with the current investigations, would be a simulation of a
nonholonomic model with a more physical constitutive equation than the
one used in section~\ref{sec:NH}.  

\section{Appendix: the nonholonomic Noether theorem}

In the derivation of the Euler-Lagrange equations, both in the free
as in the constrained case, we have restricted our attention to
\emph{vertical} variations.  While it is well known that arbitrary
variations do not yield any new information beyond the Euler-Lagrange
equations (see \cite{MPSW01}), the situation is not at all clear for
nonholonomic field theories.

This is especially important for the derivation of the nonholonomic
Noether theorem.  Therefore, we propose the following modified bundle
of reaction forces: if $F = \dual{A^\alpha_a \d y^a}$, then  
\begin{equation} \label{bundleF}
  \bar{F} := \dual{A^\alpha_a (\d y^a - y^a_\mu \d x^\mu) \wedge \d^n
    x_0} \subset \Omega^{(n + 1)}(J^1\pi).  
\end{equation}
This situation is reminiscent of the comparison between Bridges' $(n +
1)$ multisymplectic $1$-forms and the Cartan $(n + 1)$-form $\Theta_L$
in the work of Marsden and Shkoller \cite{MarsdenShkoller}, and is
similar to the variational derivation of the Cartan form: if only
vertical variations are taken into account, then the Cartan form is
missing the $\d^{n + 1}x$ term (see \cite{MPSW01}).

The precise form of the bundle $\bar{F}$ can be derived by using
arguments from Cauchy analysis (see \cite{gimmsyII}).  The idea is to
reformulate the field equations as a mechanical system on an infinite
dimensional configuration space.  The reaction forces can then be
introduced in a straightforward way on this infinite-dimensional
space, and by returning to the jet bundle one then obtains
(\ref{bundleF}).  The details of that derivation would lead us too
far; more information on this technique will appear in a forthcoming
publication (see also \cite{MyPhd}).  For now, we will simply accept
the bundle $\bar{F}$ as given.

In \cite{Vankerschaver05} a similar type of bundle was used in the
derivation of the nonholonomic momentum lemma.  From that paper, we
cite the following \emph{nonholonomic Noether theorem}:

\begin{prop} \label{NHnoether} Let $L$ be a $G$-invariant
  Lagrangian density.  Assume that $\xi \in \gothg$ is such that
  $\xi_{J^1\pi} \contraction \alpha = 0$ along $\calC$ for all $\alpha
  \in \bar{F}$.  Then the following conservation law holds:
  \[
    \d [ (j^1\phi)^\ast J^L_\xi ] = 0,
  \]
  for all sections $\phi$ of $\pi$ that are solutions of the
  nonholonomic field equations.
\end{prop}

Note that a vertical vector $v$ belongs to $F^\circ$ if and only if
$\dual{v , \alpha} = 0$ for all $\alpha \in \bar{F}$.  Only for
non-vertical vectors there is a difference between $F$ and $\bar{F}$.
Therefore, if $\xi_{J^1\pi}$ in proposition~\ref{NHnoether} is
vertical, then the nonholonomic Noether theorem follows from the
Euler-Lagrange equations (\ref{thm:EL}).  In the other case, the
techniques from \cite{Vankerschaver05} have to be used.


\ack Financial support from the Research Association--Flanders
(FWO-Vlaanderen) is gratefully acknowledged.  I would like to thank
Frans Cantrijn for useful discussions and a critical reading of this
manuscript, as well as Thomas Bridges, Manuel de Le\'on, Marcelo
Epstein, and David Mart\'{\i}n de Diego for interesting discussions
and many useful remarks.  This gratitude extends also to W. Tulczyjew
and D. Zenkov, for formulating a number of critical remarks that led
to the improved version of the nonholonomic example presented here.
Furthermore, I wish to thank an anonymous referee for pointing out an
inconsistency in an earlier version of this paper.

\end{document}